\documentclass[%
 aip,
 amsmath,amssymb,
 reprint,%
]{revtex4-1}

\usepackage{graphicx}
\usepackage{dcolumn}
\usepackage{bm}

\usepackage[utf8]{inputenc}
\usepackage[T1]{fontenc}
\usepackage{mathptmx}
\usepackage{etoolbox}

\usepackage{amsmath}
\usepackage{url}
\usepackage{array,tabularx}
\usepackage{graphics,color} 
\usepackage{epsfig} 
\usepackage{mathptmx} 
\usepackage{times} 
\usepackage{amssymb}  
\usepackage{xfrac}
\usepackage{comment}
\usepackage{gensymb}
\usepackage[export]{adjustbox}

\makeatletter
\def\@email#1#2{%
 \endgroup
 \patchcmd{\titleblock@produce}
  {\frontmatter@RRAPformat}
  {\frontmatter@RRAPformat{\produce@RRAP{*#1\href{mailto:#2}{#2}}}\frontmatter@RRAPformat}
  {}{}
}%
\makeatother
\begin{document}

\preprint{AIP/123-QED}

\title[]{Active Control Approach to Temporal Acoustic Cloaking}
\author{Or Lasri}
\author{Lea Sirota}%
 \email{leabeilkin@tauex.tau.ac.il}
\affiliation{ 
School of Mechanical Engineering, Tel Aviv University, Tel Aviv 69978, Israel
}%


\date{\today}

\begin{abstract}
We propose a realization of a transformation-based acoustic temporal cloak using an active closed-loop control approach to an equivalent electromagnetic problem. 
Unlike the more common spatial cloaks the goal of which is hiding fixed objects from detection, the goal of the temporal cloak is hiding the occurrence of events during a finite period of time. 
In electromagnetic systems, in which events represent, for example, leakage of signals from transmission lines or optical fibers, temporal cloaking solutions usually rely on nonlinear phenomena related to the fibers properties, or on modulating the properties of the propagation medium itself. 
In particular, the transformation-based solution requires modulating the constitutive parameters of the medium both in space and time.
In acoustic systems, an event may represent an object crossing a propagation channel and temporarily blocking it.
Our control approach is fully linear, where the required change in the medium parameters is programmed into the controllers and created by external actuators in real-time. This cloaking system keeps the physical medium unchanged, and enables to reprogram the cloaking parameters upon request.
We demonstrate our solution in a simulation of a one-dimensional water channel.
\end{abstract}

\maketitle

Cloaking is usually associated with hiding objects in space from electromagnetic or acoustic detection \cite{schurig2006metamaterial,cummer2007one,li2008hiding,valentine2009optical,alu2009mantle,popa2011experimental,zhu2013one,zigoneanu2014three,fan2020reconfigurable}. In one of the common approaches, which was based on transformation field theories \cite{chen2010transformation,zhang2019transformation}, spatial cloaking suggested a smooth guidance of waves around the object in a designated surrounding region, without backscattering, and with a perfectly restored arrival to the receiver \cite{schurig2006metamaterial,cummer2007one}. Such a `hole' in space was achieved by spatially manipulating the dispersion in the cloaked region, in particular by creating inhomogeneous and anisotropic constitutive parameters of the material therein. 

Alternatively, some applications may require hiding the occurrence of events of finite duration, rather than objects. 
For a medium with propagation in one dimension, such as a transmission line or an optical fiber for electromagnetic systems, an event can represent a finite-duration leakage of signal out of the fiber. 
In acoustic systems a dominantly one-dimensional propagation can be obtained inside a channel filled, e.g., with air or water. An event can then represent a finite-duration change of the channel cross-section geometry, caused, for instance, by an object or a vessel passing through it. 

Several realizations of a temporal cloak have been proposed for electromagnetic or optical systems, and carried out using the nonlinear properties of optical fibers and optical time lenses \cite{fridman2012demonstration,lukens2013temporal,chremmos2014temporal,zhou2017temporal,li2017extended,zhou2019field}. 
Another realization invoked the transformation field theory for one space and one time dimension instead of two space dimensions that are employed in the spatial transformation cloak \cite{mccall2010spacetime,kinsler2014cloaks}.   
This was proposed to be achieved by manipulating the constitutive parameters of the medium in space and time, thereby creating an analogous hole in space-time. 
Since in this situation geometrical deflection and reassembly of the field was not required, the underlying principle was slowing down the detection field in the cloaked region until its complete suppression for a limited time period during which the event occurred, and then speeding up the field to compensate for the delay. This manipulation was achieved without reflection neither to the emitter nor to the receiver.

The goal of this work is to design a realization of acoustic temporal cloak, which will be fully linear and will not require to change the properties of the actual propagation medium. 
We consider the transformation-based approach in \cite{mccall2010spacetime,kinsler2014cloaks}, and propose an alternative realization of the cloak as an active control problem, which is applied to an otherwise uniform medium. 
In this realization, the required space and time dependent properties, which are effective mass density and bulk modulus in the cloaked region, as well as additional cross-coupling terms of pressure and flow velocity, are created by the controllers in a real-time measurement-based feedback operation. 
The cloak parameters then become reconfigurable, as they can be reprogrammed by the user into the controller.

Actively-controlled measurement-based design has recently emerged in diverse electromagnetic, acoustic and elastic applications \cite{akl2010multi,popa2015active,borsing2019cloaking,hofmann2019chiral,sirota2019tunable,sirota2019active,sirota2020modeling,becker2020real,scheibner2020non,rosa2020dynamics,cho2020digitally,ghatak2020observation,kotwal2021active,you2021reprogrammable,geib2021tunable,li2021active,stojanoska2022non}, enabling exotic wave dynamics, such as non-reciprocal propagation, adaptive refocusing, in-domain absorbers, or artificial boundary conditions for simulation domain scaling. 
Here we utilize this approach for temporal cloaking in a one-dimensional acoustic medium.

\begin{figure}[htpb] 
\begin{center}
 \includegraphics[height=3.4 cm]{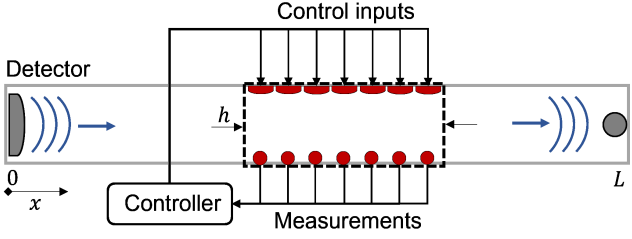}
\end{center}
 \caption{Active closed-loop control realization of the transformation based acoustic temporal cloak of Refs. \cite{mccall2010spacetime,kinsler2014cloaks} in a water channel. A control frame (dashed-black) of length $h$, mounted at the channel wall, comprises an array of monopole actuators (red disks) and pressure sensors (red circles), facing inward. The actuators, based on real-time measurements of the sensors, create the required acoustic pressure field in the channel, using the control algorithm that is programmed into the controller. As a result, a null intensity window is opened in space-time, undetected by a detection system at the channel ends (grey disk and circle).
 }
\label{T_cloak_ours}
\end{figure}

Our representative system is a water channel, comprising an emitter-receiver pair at one end, $x=0$, and another receiver at the other end, $x=L$, as illustrated in Fig. \ref{T_cloak_ours}. These emitter and receivers (grey elements) constitute a detection system, the role of which is to intercept a disturbance in the channel, where the emitter produces a signal continuously. 
It is then assumed that the channel cross-section within the region outlined by the dashed black frame can be narrowed and even blocked completely for the time duration $t_1<t<t_2$, imitating, for example, an underwater vessel passing across the channel. 

To keep this event undetected, we aim at generating an active-control-based temporal cloak along the entire frame, which we thereby denote by the cloaking region. This control frame of length $h$ is mounted at the channel wall and can be removed at will. An array of control emitters and receivers (red elements) are spread along the frame, facing inwards.

We now describe the target closed-loop cloaking system and the corresponding control design that creates the required reconfigurable medium. 
The space-time cloaking mechanism itself was originally designed for an electromagnetic medium, which is the target system of our controller to create in closed-loop in an acoustic medium. 

\begin{figure}[htpb]
\begin{center}
\includegraphics[height=6.4cm]{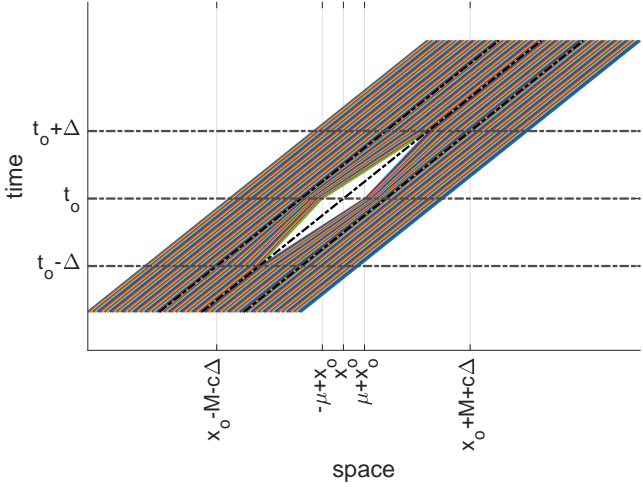}
   \end{center}
\caption{Ray trajectories in the original domain $(x,t)$. The cloaking action takes place in a parallelogram of spatial extension determined by $M$, and a resulting temporal extension determined by $\Delta$ and wave velocity $c$. The null intensity window of maximal width $\mu$ (white internal parallelogram) is opened due to the field distortion around the event occurrence at $(x_0,t_0)$, based on the transformation \eqref{eq:AB}-\eqref{eq:theta_eta}.}
\label{fig:cloak_scheme}
\end{figure}

The cloaking region in space-time was suggested as a parallelogram \cite{mccall2010spacetime,kinsler2014cloaks,cornelius2014finite}, as depicted by the dashed black lines in Fig. \ref{fig:cloak_scheme}. The length of the sides parallel to the $x$ axis is defined by $2M$. The length of the total $x$ axis projection, centered at $x_0$, is given by $h$, which is the length of the control frame. The $t$ axis projection, i.e. the cloaking time window, is of length $t_2-t_1=2\Delta$, where $t_1=t_0-\Delta$ and $t_2=t_0+\Delta$ are the beginning and end times, respectively.
The temporal opening is a resulting parameter dependent on the spatial opening through the wave velocity $c$, hence for a longer duration $\Delta$ one has to increase $M$. 
The cloaking result, which is the state of intensity null of the detection wave, the dark region, or the hole in space-time, is obtained in the internal parallelogram (white). This is the region where an event, occurring during $t\in(t_1,t_2)$, remains undetected. The event itself, however, has no effect on the cloak opening and closing process. The point $(x_0,t_0)$ indicates the cloak center; $t_0$ is the time when the null intensity created by the cloak has the widest spatial support, $2\mu$. 
The cloaking operation exists throughout the external parallelogram. 

At $t<t_0$ the cloaking action includes slowing down and speeding up the detection field at the bottom-left and bottom-right quarters, respectively, and switched at $t>t_0$ to speeding up and slowing down at the top-left and top-right quarters. 
Outside the external parallelogram the field remains unperturbed, as represented in Fig. \ref{fig:cloak_scheme} by rays of slope $1/c$. 
The slowing down and speeding up regions are respectively represented by rays of a slope higher and lower than $1/c$. 
Inside the parallelogram the cloak was originally described by four abstract fields $A(x,t),B(x,t),a(x,t),b(x,t)$, coupled by two differential constitutive relations and two algebraic relations,
\begin{subequations}   \label{eq:AB}
\begin{align}
    A_t&=b_x, \quad a=\theta(t)\left[\alpha_0A+\eta(x,t)B\right], \\
    B_t&=a_x, \quad b=\theta(t)\left[\eta(x,t)A+\gamma_0B\right].
\end{align}
\end{subequations}
Here, $\alpha_0$ and $\gamma_0$ are the medium parameters, which are constant when the medium is uniform, i.e. outside the external parallelogram. Subscripts symbolize partial derivatives. $\theta(t)$ and $\eta(x,t)$ are the cloak parameters, which are functions of space and time.
Outside of the cloaking region, as well as inside it when the control system creating the cloak is turned off, we obtain $\theta(t)=1$ and $\eta(x,t)=0$. The four-field cross-coupling then breaks, reducing to a two-field coupling in \eqref{eq:AB}, with $a=\alpha_0A$ and $b=\gamma_0B$. 
Inside the cloaking region, $\theta(t)$ and $\eta(x,t)$ are obtained from the transformation
\begin{equation}   \label{eq:theta_eta}
 \begin{pmatrix} \tilde{} & \theta^{-1}(t) \\ \tilde{} & -\eta(x,t) \end{pmatrix}=
J^{-1}\begin{pmatrix} 0 & 1 \\ \alpha_0\gamma_0 & 0 \end{pmatrix}J, \quad J=\left(
    \begin{array}{cc}
      t'_t   &  t'_x \\
       x'_t   &  x'_x 
    \end{array}\right),
\end{equation}
where $J$ is the Jacobian matrix, and $\tilde{}$ symbolizes redundant information. 
The coordinates $(x',t')$ describe a transformed domain, in which the wave propagation in the cloaking region appears undistorted. These coordinates are aligned with the required distortion pattern in the physical $(x,t)$ domain, governed by \eqref{eq:AB}-\eqref{eq:theta_eta}. The exact mapping from $(x',t')$ to $(x,t)$ depends both on the medium properties through $c$ and the cloak properties through $M$, $\mu$, $t_0$ and $x_0$ \cite{kinsler2014cloaks,cornelius2014finite}, and is given by
\begin{subequations} \label{eq:coords}
    \begin{align}
    t'&=t, \\
    x'&=\left(x-X\right)\left(1-\frac{\mu}{M}\frac{t-T}{t_0-T}\right)^{-1}+X, \\
    T&=t_0+\Delta \cdot sign(t-t_0), \\
    X&=\frac{x_0}{M}+\frac{c}{M}(t-t_0)+sign(x-x_0).
    \end{align}
\end{subequations}
Applying the transformation \eqref{eq:theta_eta} with the coordinates \eqref{eq:coords}, one obtains the cloaking parameters
\begin{subequations} \label{eq:cloaking_parameters}
    \begin{align}
    \theta(t)&=\left(1-\frac{\mu}{M}\frac{t-T}{t_0-T}\right)^{-1}, \label{eq:cloak_param_theta} \\
    \eta(x,t)&=\theta^2(t)\left[\left(\frac{\mu}{M}-1\right)c+\frac{M}{\mu}\frac{x-x_0\mp M}{t_0-T}\right]+c. \label{eq:cloak_param_eta}
    \end{align}
\end{subequations}

\begin{figure}
    \centering
    \includegraphics[height=6.7cm]{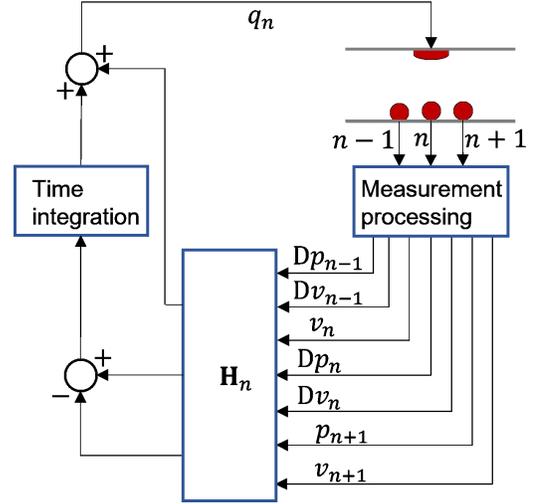}
    \caption{Control algorithm implementation. The measurements at the $n$ site and the adjacent sites are processed by the controller. The controller translates them into approximated spatial derivatives of pressure and velocity defined by the operator D, multiplies by the field coupling parameters in the matrix $\textbf{H}_n$ in \eqref{eq:controller}, integrates in time as defined in \eqref{eq:control}, and sums up to the total monopole input $q_n$.}
    \label{fig:unit_cell}
\end{figure}

Having the abstract field formulation of the target cloak \eqref{eq:AB}-\eqref{eq:theta_eta} at hand, we are ready for the main part of this work, which is mapping the formulation to an acoustic medium and designing the control system that will reach this target in real-time.
We consider, for example, the mapping
\begin{equation}   \label{eq:map}
    A=v, \quad B=p, \quad \alpha_0=-b_0, \quad \gamma_0=-1/\rho_0,
\end{equation}
where $p(x,t)$ is the sound pressure field, $v(x,t)$ is the flow velocity field, $b_0$ is the bulk modulus and $\rho_0$ is the mass density of the acoustic medium.
We note that the mapping in \eqref{eq:map} is not unique. A different mapping will lead to a different control strategy, but the final result of intensity null in the internal parallelogram has to be the same.
Substituting the fields \eqref{eq:map} and the cloaking parameters \eqref{eq:theta_eta} in \eqref{eq:AB}, and omitting the $x$ and $t$ dependence of $p$, $v$, $\theta$ and $\eta$ for brevity, we obtain
\begin{subequations}   \label{eq:target}
    \begin{align}
        p_t&=\theta\left(-b_0v_x+\eta_x p+\eta p_x\right) \label{eq:CL_p}, \\
        v_t&=\theta\left(-\rho_0^{-1}p_x+\eta_x v+\eta v_x\right). \label{eq:CL_v}
    \end{align}
\end{subequations}
The system in \eqref{eq:target} is the target system that the control inputs in Fig. \ref{T_cloak_ours} need to generate in closed loop in a uniform medium governed by the free field wave equation. 
As these inputs are planned to be operated through the channel wall and not through its cross-section, we assume all the actuators to be of monopole type.
Denoting these inputs by $q(x,t)$, we set
\begin{equation}     \label{eq:control}
\begin{split}
    q&=(1-\theta)v_x+\theta b_0^{-1}\eta_xp+\theta\eta b_0^{-1}p_x \\
    &+\int_0^t\left[(1-\theta)p_{xx}+\rho_0\theta\left(2\eta_xv_x+\eta v_{xx}\right)\right]dt.
    \end{split}
\end{equation}
The control algorithm in \eqref{eq:control} is given in a spatially continuous form, whereas the actual actuators are spatially discrete and spaced at a nonzero distance, which is determined by their physical size. 
The realization of \eqref{eq:control} for the site $n$ actuator is illustrated in Fig. \ref{fig:unit_cell}. 
The pressure field measurements at the site $n$ and the adjacent sites $n+1$ and $n-1$, as well as their differences normalized by the spacing, both for the pressure $p$ and the velocity $v$ (where $v$ itself is approximated as a scaled pressure gradient) are fed into the control matrix $\textbf{H}_n$, given by 
\begin{equation}     \label{eq:controller}
\def\arraystretch{1.25}
\textbf{H}_n=\left(\begin{array}{ccc}
    0 & 0 & 1-\theta_n \\
    0 & 0 & \rho_0\theta_n\eta_n \\
    0 & 0 & \rho_0\theta_n\eta_{x\_n} \\
    b_0^{-1}\theta_n\eta_{n+1} & 1-\theta_n & 0 \\
    1-\theta_n & \rho_0\theta_n\eta_{n+1} & 0 \\
    b_0^{-1}\theta_n\eta_{x\_{n+1}} & 0 & 0 \\
    0 & \rho_0\theta_n\eta_{x\_{n+1}} & 0
\end{array}\right)^T.
\end{equation}
The control matrix includes the cloaking parameters $\theta$, $\eta$ and $\eta_x$, calculated at the $n$ and $n+1$ sites. The parameter $\theta$ is independent of space and thus can be calculated at any site.
Finally, numerical time integration is applied to the second and third controller's outputs difference, which is then summed with the first one to produce the total control input $q_n$. 

\begin{figure*}[htpb]
    \centering
    \setlength{\tabcolsep}{-1pt}
    \renewcommand{\arraystretch}{1.2}
    \begin{tabular}{ccc}
    \textbf{(a)} & \textbf{(c)} & \textbf{(e)} \\
    \includegraphics[width=6.0cm, valign=c]{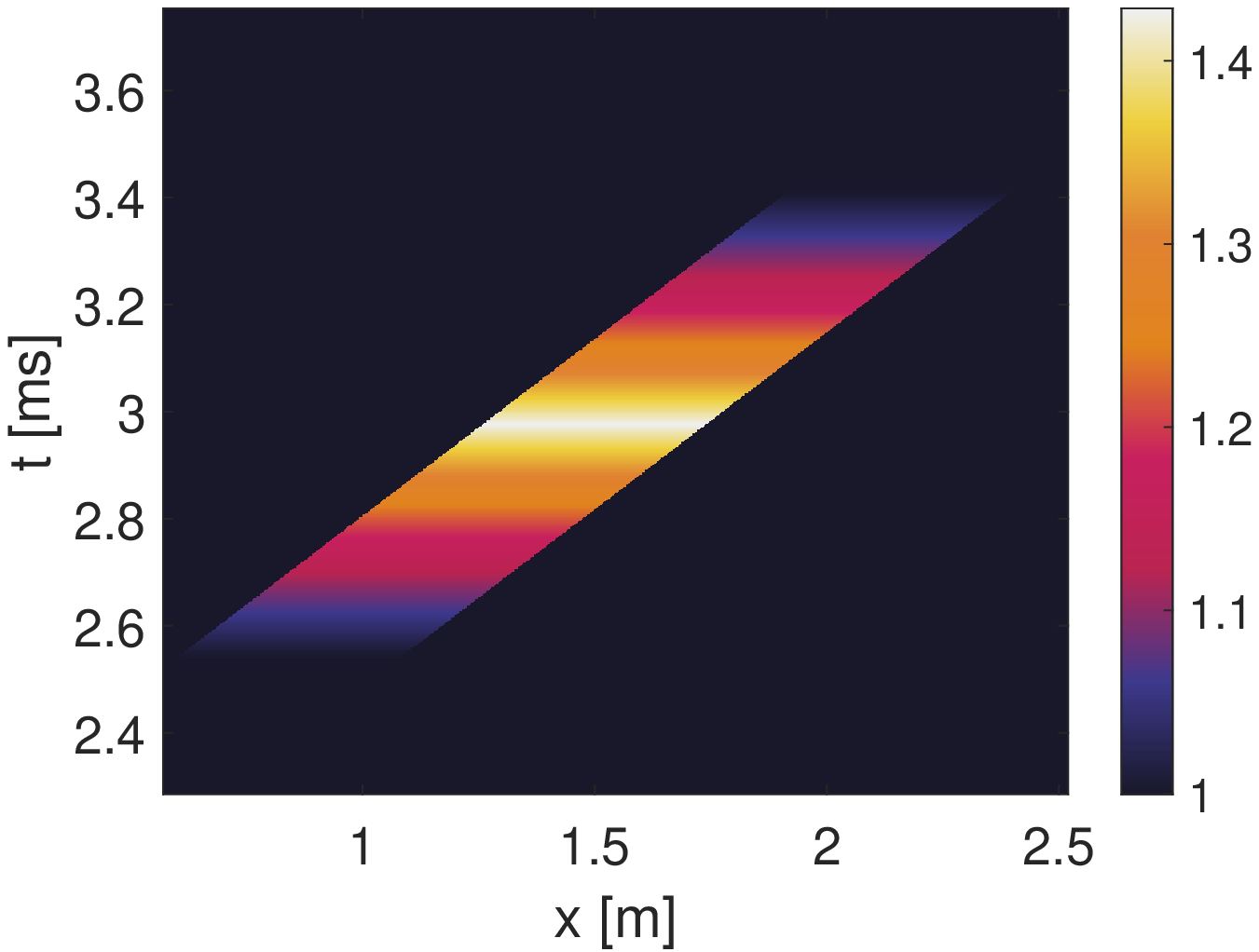} & \includegraphics[width=6.0cm, valign=c]{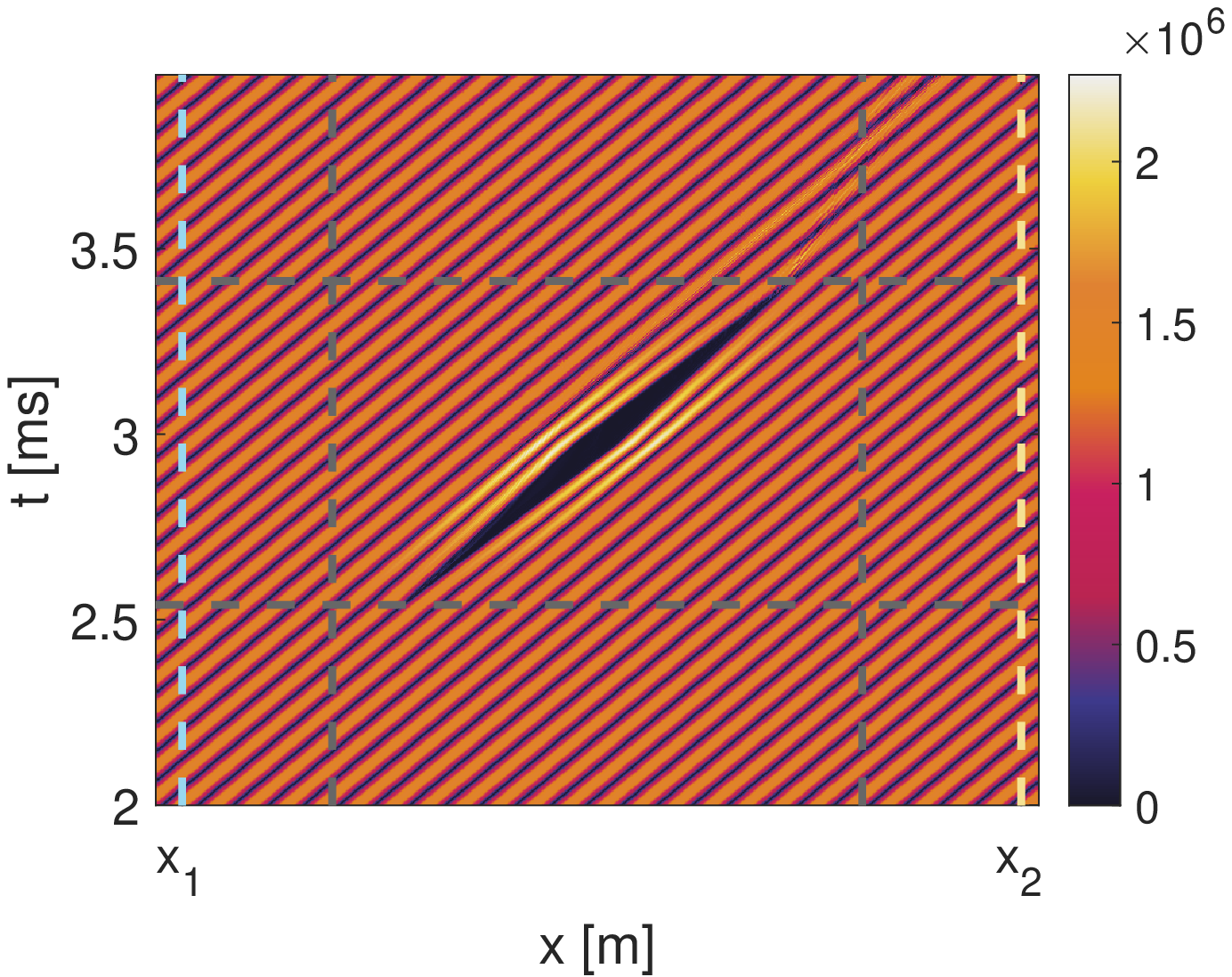} &  \includegraphics[width=6.0cm, valign=c]{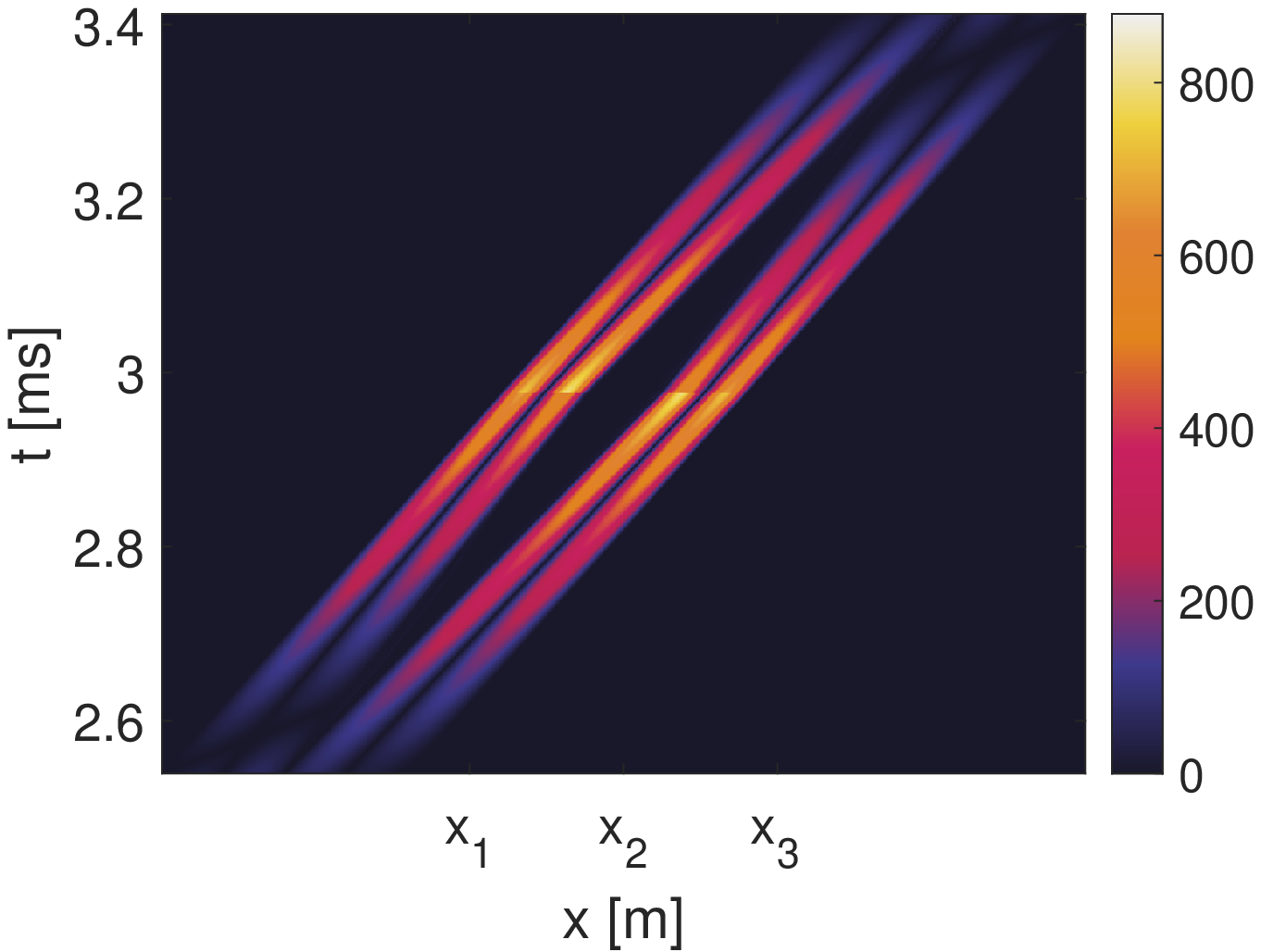} \\
    \textbf{(b)} & \textbf{(d)} & \textbf{(f)} \\    
          \includegraphics[width=6.0cm, valign=c]{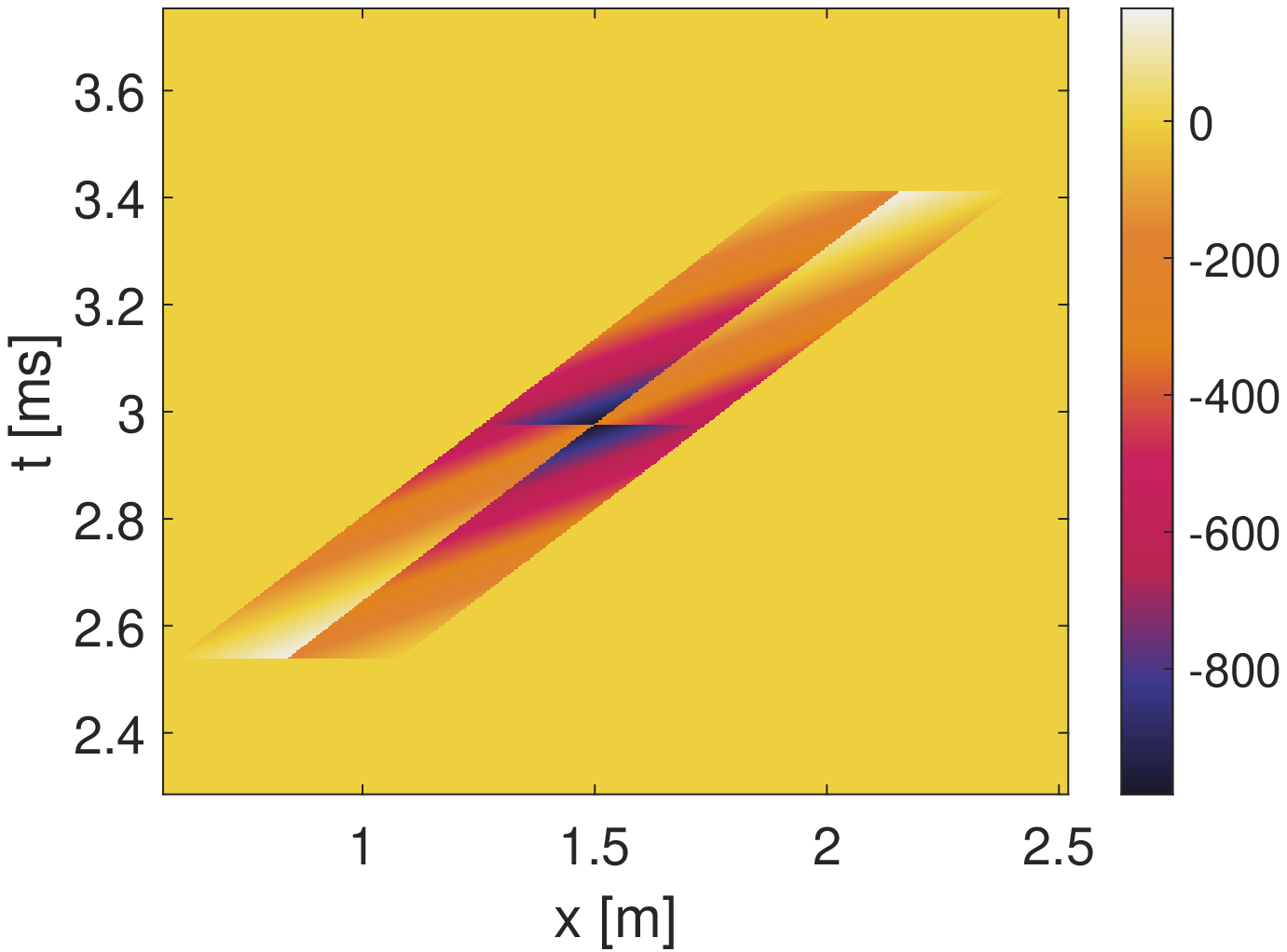} & \includegraphics[width=6.0cm, valign=c]{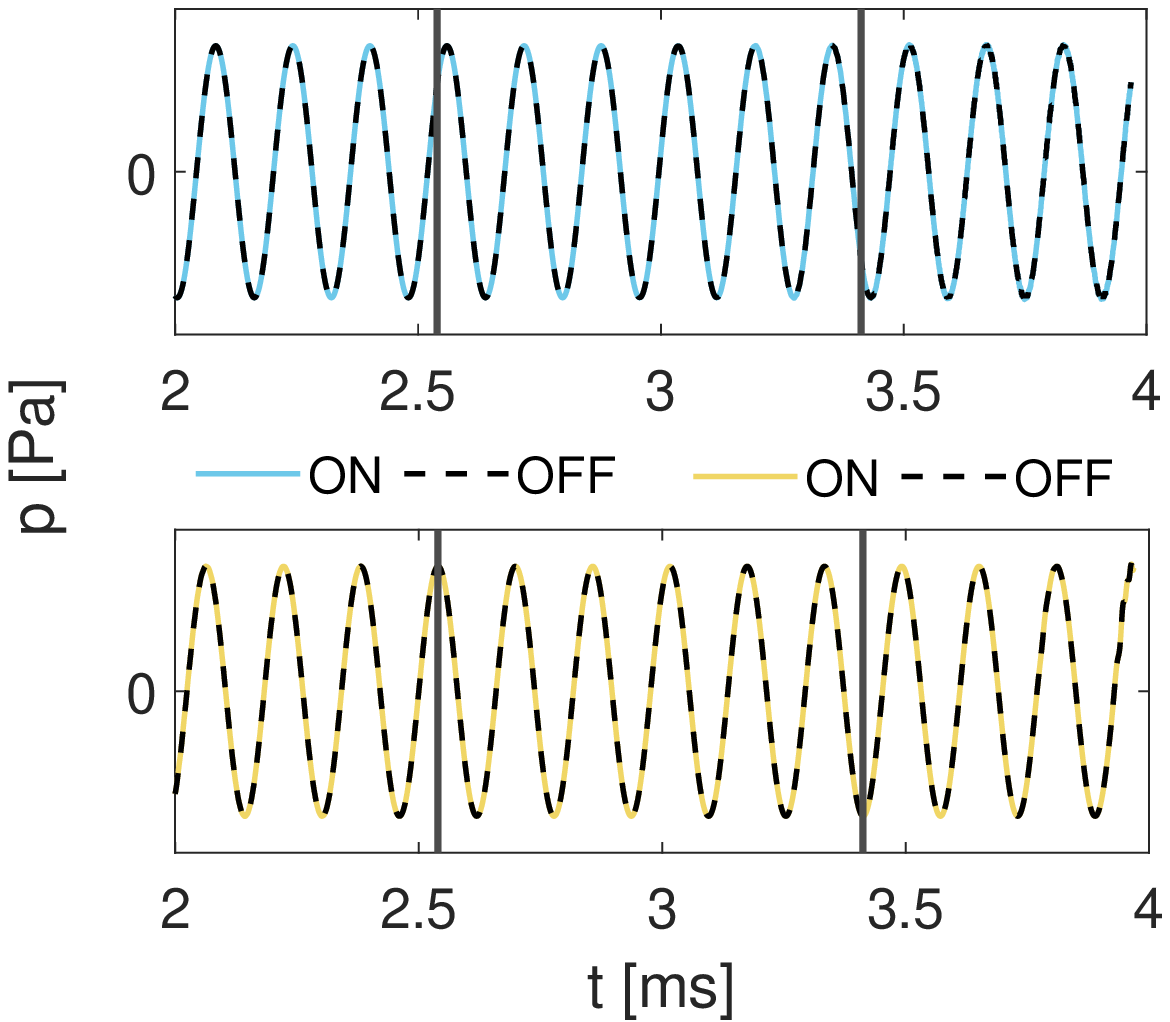} &  \includegraphics[width=6.0cm, valign=c]{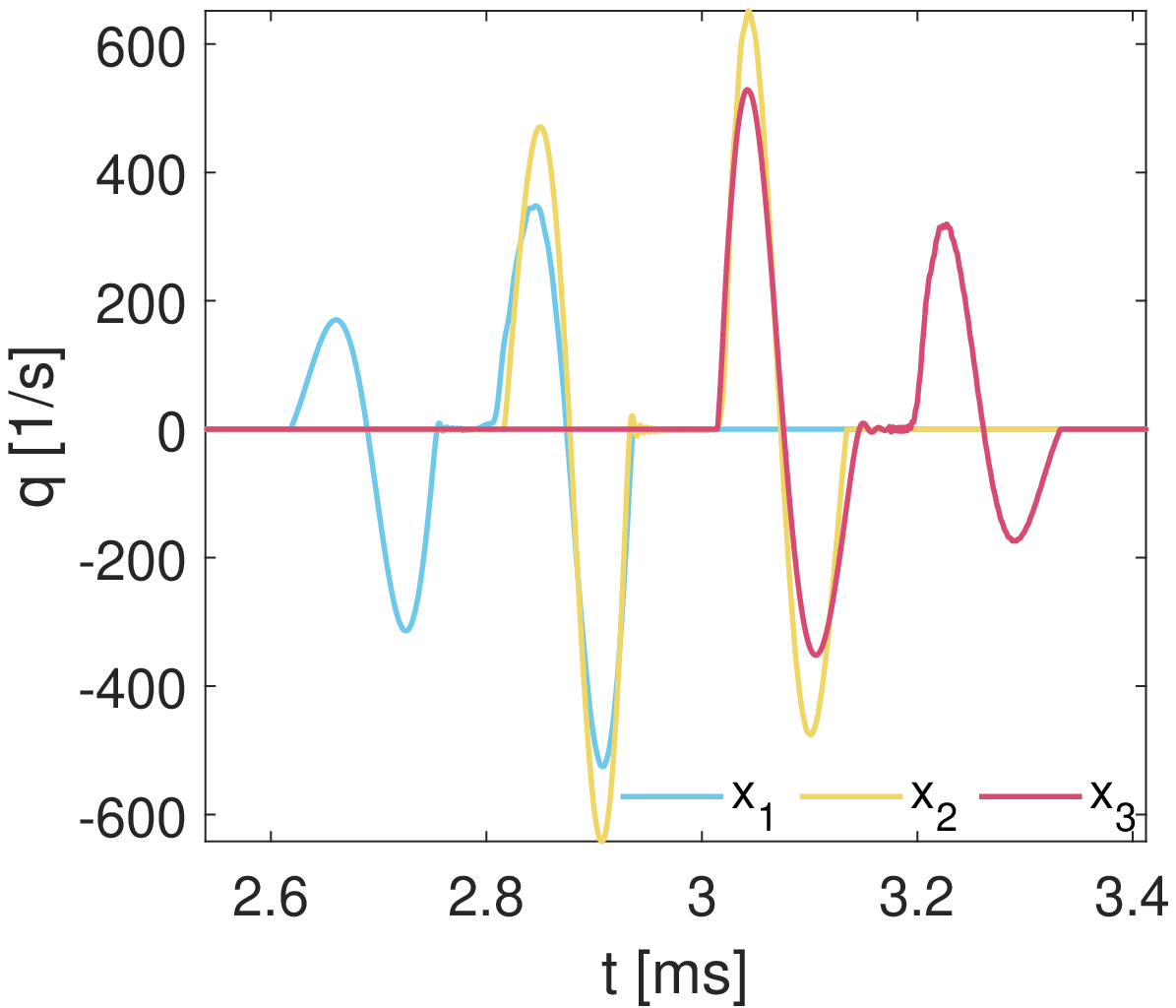}
    \end{tabular}
    \caption{Performance demonstration of the control-based temporal cloak of opening $M=0.24$ m and $\mu=0.072$ m in a water channel of length $L=3$ m. (a), (b) - the field coupling parameters $\theta(t)$ and $\eta(x,t)$, respectively, calculated from \eqref{eq:theta_eta}. (c) - The pressure field space-time distribution created in real-time by the controller \eqref{eq:control}-\eqref{eq:controller}, here simulated via the finite difference time domain method. The null intensity region is opened and closed as expected. The locations $x_1$ and $x_2$ indicate the detection observers. (d) - The pressure field time responses at the $x_1$, top, and $x_2$, bottom, when the control is on (solid), indicating the cloak action, and off (dashed), indicating free field propagation. The responses are nearly identical. (e) - The space-time distribution of the control signal $q$, defined in \eqref{eq:control}-\eqref{eq:controller}. The control action takes place in the external parallelogram region only, in which the field distortion is required. (f) - The time domain response of the control signals at locations $x_1$ (blue), $x_2$ (yellow) and $x_3$ (red), as indicated on top of panel (e), respectively representing the actuators behind, at and in front of the event occurrence. }
    \label{fig:simulations}
\end{figure*}

We now demonstrate the performance of the temporal cloak realized by the control algorithm in \eqref{eq:control}-\eqref{eq:controller}. The validation was carried out via a finite difference time domain simulation of a water channel of length $L=3$ m and medium parameters $\rho_0=1024$ kg/m$^3$, $b_0=2.34\cdot 10^9$ Pa, leading to the wave velocity $c=1512$ m/s.
The spatial and temporal steps were taken as $dx=10^{-3}$ m and $dt=5.3\cdot 10^{-7}$ s, respectively. 
The size of the control frame was set to $h=1.8$ m. The parameter $h$ determines the maximal spatial extension of the cloak, and thus the maximal possible $M$. The actual $M$, however, can be made smaller by activating only part of the control actuators along the frame, which is done automatically by our control program when the desired parameter $M$ is fed in. 
Here, $M$ was set to 0.24 m, implying the time window $2\Delta=0.004$ s. The inner spatial opening was set to $\mu=0.072$ m. 
The resulting cloaking parameters $\theta(t)$ and $\eta(x,t)$ in \eqref{eq:cloak_param_theta}-\eqref{eq:cloak_param_eta} were calculated by the program, and are depicted in Fig. \ref{fig:simulations}(a)-(b). 
The simulations were carried out using step-wise update of the feedback law in \eqref{eq:control}-\eqref{eq:controller}.

For a plane interception wave of frequency $f_0=6.3$ kHz and the corresponding wavelength $\lambda=24$ cm, the controller generated the sound pressure field that is depicted in Fig. \ref{fig:simulations}(c).
We observe a dark parallelogram, which indicates the required hole in space-time. In this hole the interception field is missing, allowing undetected trespassing, as indeed confirmed by the observers responses. It should be noted, however, that since the coordinate transformation \eqref{eq:theta_eta}-\eqref{eq:cloaking_parameters} is frequency independent, any excitation waveform is valid. The plane wave was chosen since it exists at all times, emphasizing better the hole in space-time. The pressure field responses for the observer to the left of the cloaking region and the one to the right is depicted in Fig. \ref{fig:simulations}(d), top and bottom subplots, respectively. 
These responses were obtained for actuators and sensors of 2 cm in diameter, implying the total of 90 actuator-sensor pairs placed along the frame.
For both observers, the responses when control is on (solid curve), i.e. the cloak is created, and off (dashed curve), i.e. free field, are nearly indistinguishable. 
This confirms that the control law \eqref{eq:control}-\eqref{eq:controller} generates the required transformation without introducing any back-scattering at the cloak opening or closing stage.

\begin{figure}[htpb]
    \centering
    \setlength{\tabcolsep}{-6pt}
    \begin{tabular}{cc}
    \textbf{(a)} & \textbf{(b)}  \\
    \includegraphics[width=4.7cm, valign=t]{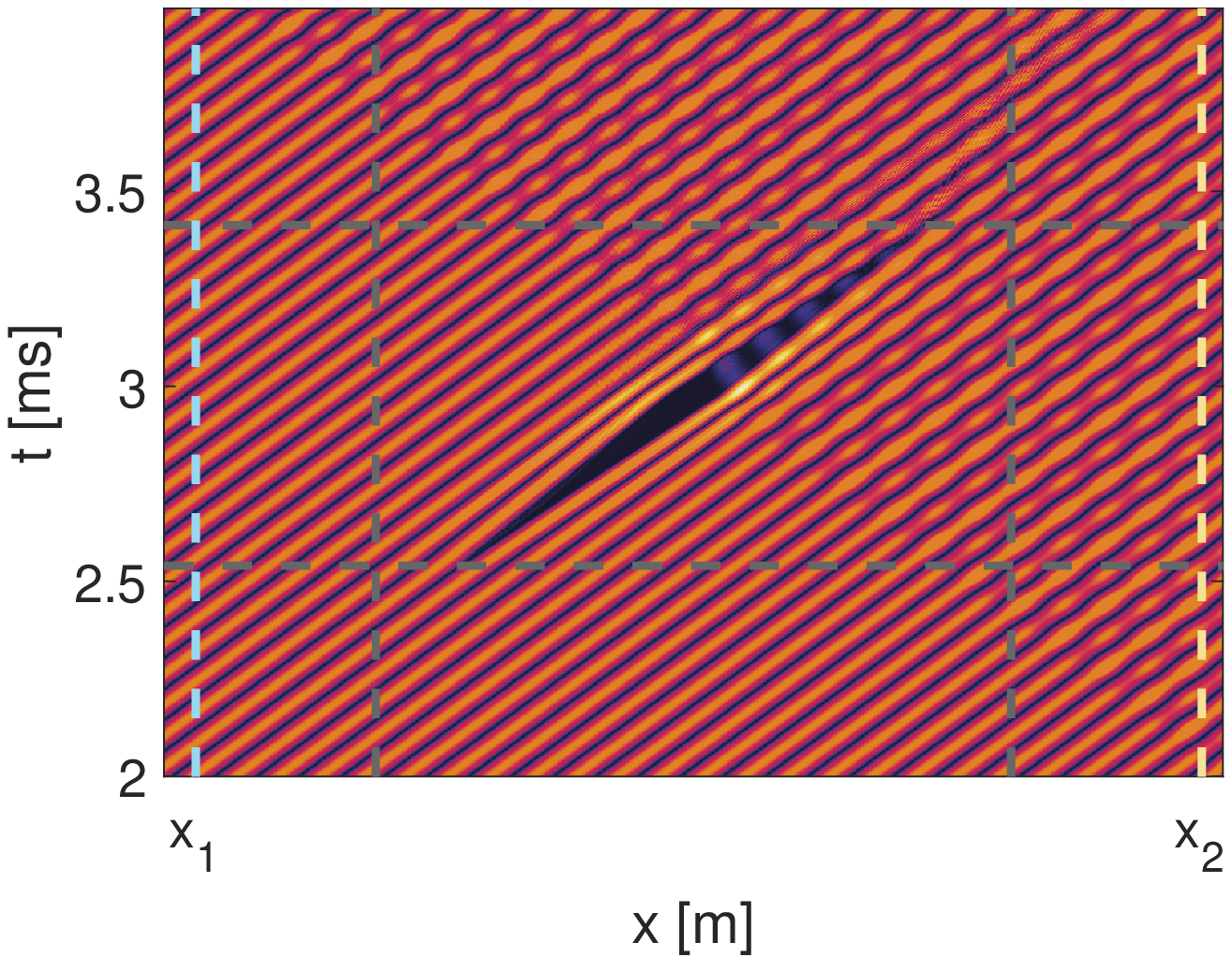} & \includegraphics[width=4.7cm, valign=t]{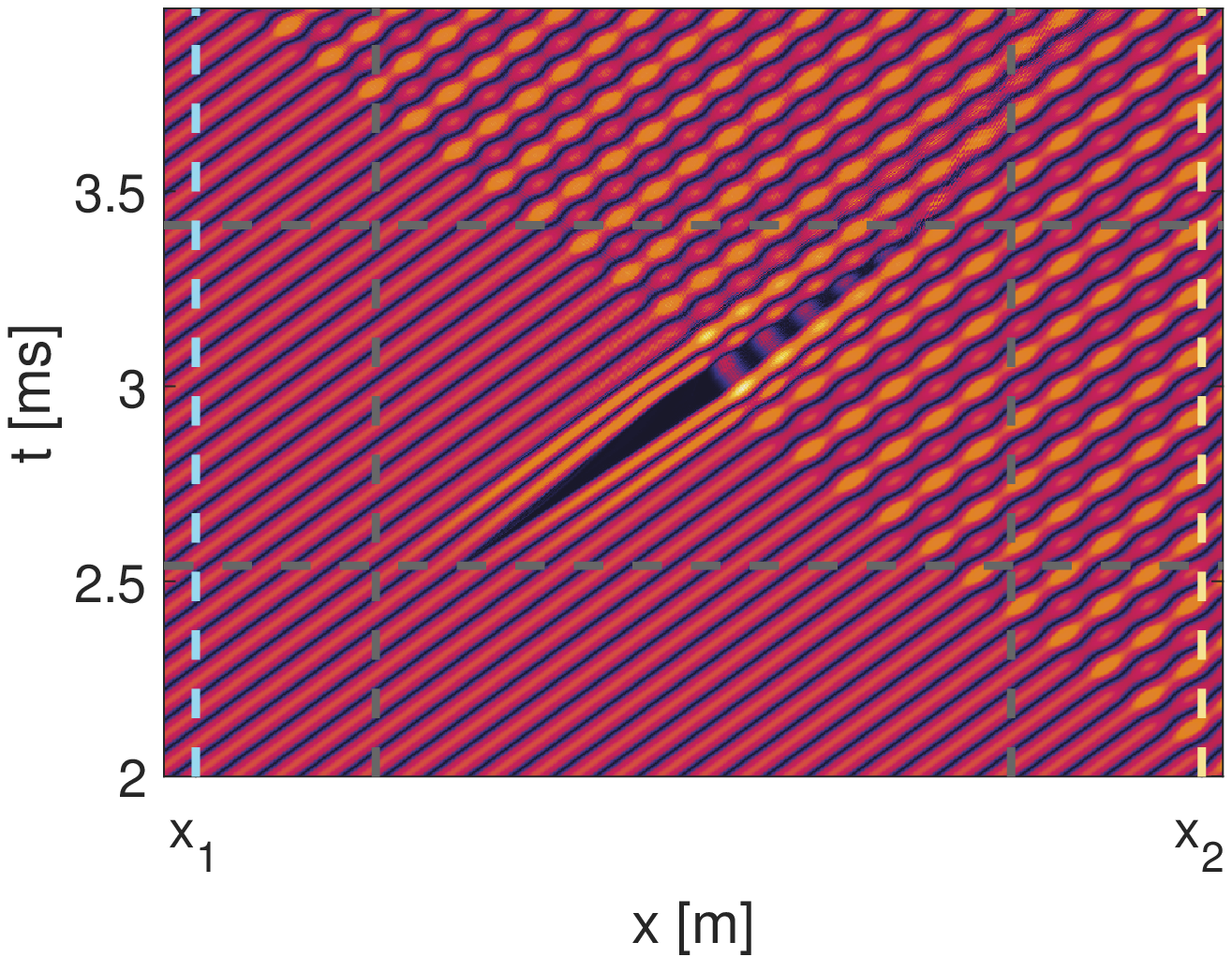} \\
    \includegraphics[width=4.5cm, valign=t]{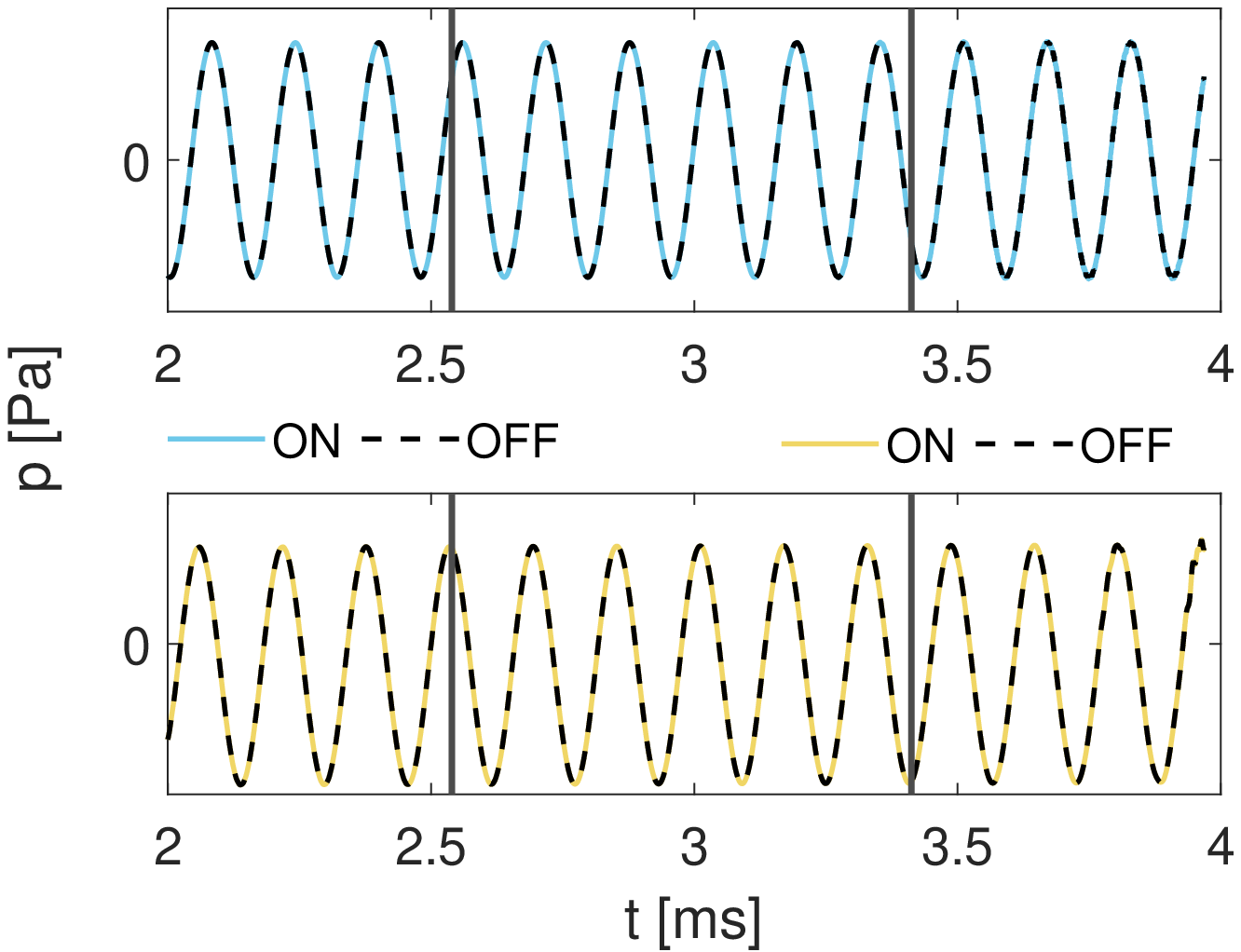} & \includegraphics[width=4.5cm, valign=t]{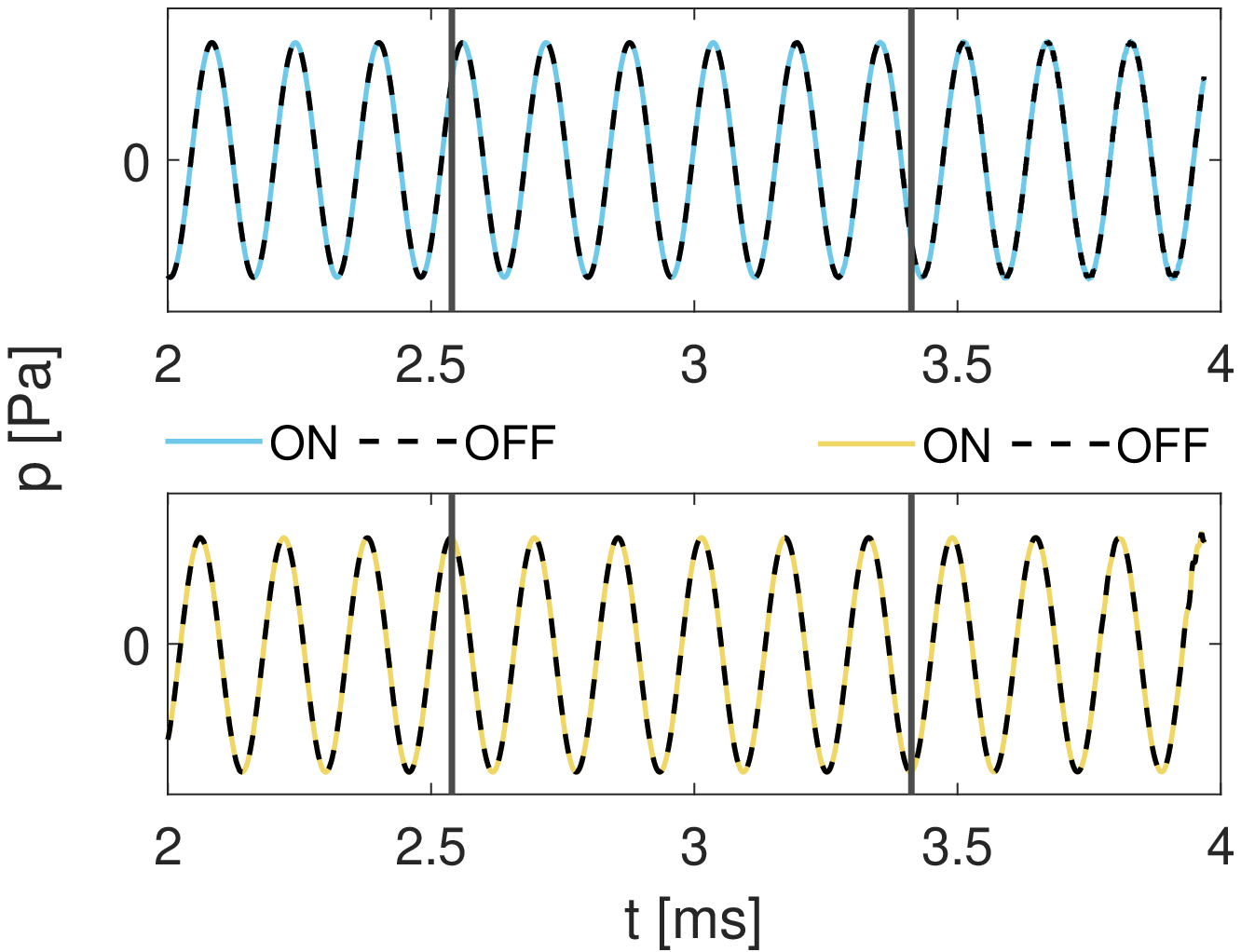}
    \end{tabular}
    \caption{Measurement error effect. Pressure field distribution (top) and the observers response (bottom) for $|0.5|\%$, (a), $|1|\%$, (b), error.}
    \label{fig:meas_err}
\end{figure}

The control inputs $q$ that created the cloaking pressure field of Fig. \ref{fig:simulations}(c) in real-time are depicted in Fig. \ref{fig:simulations}(e) as a function of space and time. 
These signals exist only during the cloaking operation, which is in the external parallelogram excluding the internal one, and obtain finite values there.
The abrupt transitions between the slow down and speed up zones, as stems from the coordinate transformation, imply that the control effort becomes higher around the transition zone. This is illustrated in Fig. \ref{fig:simulations}(f) for control signals of representative actuators behind, $x_1$, at, $x_2$, and in front, $x_3$, the event location $x_0$, plotted as a function of time. 
The smoothness and convergence of the control signals in Fig. \ref{fig:simulations}(e), as well as of the pressure field in Fig. \ref{fig:simulations}(c) , demonstrate the internal stability of the underlying dynamical system.
Finally, we test the robustness of the algorithm in \eqref{eq:control}-\eqref{eq:controller} to errors in pressure measurements. As given in Fig. \ref{fig:meas_err}, although the pressure distribution in the cloak closing region is affected, for errors up to $|1|\%$ the response at the observers remains nearly unaltered.

To conclude, we proposed an active control realization of an acoustic transformation-based temporal cloak, originally designed in Refs. \cite{mccall2010spacetime,kinsler2014cloaks} for an electromagnetic medium. 
For a representative system of a water channel the control operation was designed to be carried out by an array of monopole acoustic actuators and a complimentary array of sensors mounted in the channel wall, as depicted in Fig. \ref{T_cloak_ours}. 
The control principle is to create in real-time the pressure and velocity fields required by the transformation, Fig. \ref{fig:cloak_scheme}, on top of an otherwise uniform medium, so that free field propagation is resumed when the control is switched off. 
Our proposed algorithm is given in \eqref{eq:control} in a spatially continuous form, and illustrated in Fig. \ref{fig:unit_cell} for an actual discrete implementation.
The measurements performed by the sensors are processed into pressure, velocity, and their differences at each actuation site $n$, passed through the controller gain matrix $\textbf{H}_n$ in \eqref{eq:controller}, integrated, and fed back to the actuators.
We tested our algorithm in a numerical simulation of a 3 m long channel and a 6.3 kHz detection source wave. The controller opened the required hole in space-time with smooth and converging control signal responses, Fig. \ref{fig:simulations}. The algorithm was quite robust to measurement errors up to $|1\%|$, Fig. \ref{fig:meas_err}. Our control platform is versatile, enabling the cloaking parameters to be reprogrammed by the user.

\textit{We thank Viacheslav (Slava) Krylov and Amir Boag for insightful discussions of the results of this work. We thank Paul Kinsler and Martin McCall for insightful discussions about transformation-based temporal cloaking.}



\bibliography{aipsamp}

\begin{thebibliography}{37}%
\makeatletter
\providecommand \@ifxundefined [1]{%
 \@ifx{#1\undefined}
}%
\providecommand \@ifnum [1]{%
 \ifnum #1\expandafter \@firstoftwo
 \else \expandafter \@secondoftwo
 \fi
}%
\providecommand \@ifx [1]{%
 \ifx #1\expandafter \@firstoftwo
 \else \expandafter \@secondoftwo
 \fi
}%
\providecommand \natexlab [1]{#1}%
\providecommand \enquote  [1]{``#1''}%
\providecommand \bibnamefont  [1]{#1}%
\providecommand \bibfnamefont [1]{#1}%
\providecommand \citenamefont [1]{#1}%
\providecommand \href@noop [0]{\@secondoftwo}%
\providecommand \href [0]{\begingroup \@sanitize@url \@href}%
\providecommand \@href[1]{\@@startlink{#1}\@@href}%
\providecommand \@@href[1]{\endgroup#1\@@endlink}%
\providecommand \@sanitize@url [0]{\catcode `\\12\catcode `\$12\catcode
  `\&12\catcode `\#12\catcode `\^12\catcode `\_12\catcode `\%12\relax}%
\providecommand \@@startlink[1]{}%
\providecommand \@@endlink[0]{}%
\providecommand \url  [0]{\begingroup\@sanitize@url \@url }%
\providecommand \@url [1]{\endgroup\@href {#1}{\urlprefix }}%
\providecommand \urlprefix  [0]{URL }%
\providecommand \Eprint [0]{\href }%
\providecommand \doibase [0]{http://dx.doi.org/}%
\providecommand \selectlanguage [0]{\@gobble}%
\providecommand \bibinfo  [0]{\@secondoftwo}%
\providecommand \bibfield  [0]{\@secondoftwo}%
\providecommand \translation [1]{[#1]}%
\providecommand \BibitemOpen [0]{}%
\providecommand \bibitemStop [0]{}%
\providecommand \bibitemNoStop [0]{.\EOS\space}%
\providecommand \EOS [0]{\spacefactor3000\relax}%
\providecommand \BibitemShut  [1]{\csname bibitem#1\endcsname}%
\let\auto@bib@innerbib\@empty
\bibitem [{\citenamefont {Schurig}\ \emph {et~al.}(2006)\citenamefont
  {Schurig}, \citenamefont {Mock}, \citenamefont {Justice}, \citenamefont
  {Cummer}, \citenamefont {Pendry}, \citenamefont {Starr},\ and\ \citenamefont
  {Smith}}]{schurig2006metamaterial}%
  \BibitemOpen
  \bibfield  {author} {\bibinfo {author} {\bibfnamefont {D.}~\bibnamefont
  {Schurig}}, \bibinfo {author} {\bibfnamefont {J.~J.}\ \bibnamefont {Mock}},
  \bibinfo {author} {\bibfnamefont {B.}~\bibnamefont {Justice}}, \bibinfo
  {author} {\bibfnamefont {S.~A.}\ \bibnamefont {Cummer}}, \bibinfo {author}
  {\bibfnamefont {J.~B.}\ \bibnamefont {Pendry}}, \bibinfo {author}
  {\bibfnamefont {A.~F.}\ \bibnamefont {Starr}}, \ and\ \bibinfo {author}
  {\bibfnamefont {D.~R.}\ \bibnamefont {Smith}},\ }\bibfield  {title} {\enquote
  {\bibinfo {title} {Metamaterial electromagnetic cloak at microwave
  frequencies},}\ }\href@noop {} {\bibfield  {journal} {\bibinfo  {journal}
  {Science}\ }\textbf {\bibinfo {volume} {314}},\ \bibinfo {pages} {977--980}
  (\bibinfo {year} {2006})}\BibitemShut {NoStop}%
\bibitem [{\citenamefont {Cummer}\ and\ \citenamefont
  {Schurig}(2007)}]{cummer2007one}%
  \BibitemOpen
  \bibfield  {author} {\bibinfo {author} {\bibfnamefont {S.~A.}\ \bibnamefont
  {Cummer}}\ and\ \bibinfo {author} {\bibfnamefont {D.}~\bibnamefont
  {Schurig}},\ }\bibfield  {title} {\enquote {\bibinfo {title} {One path to
  acoustic cloaking},}\ }\href@noop {} {\bibfield  {journal} {\bibinfo
  {journal} {New Journal of Physics}\ }\textbf {\bibinfo {volume} {9}},\
  \bibinfo {pages} {45} (\bibinfo {year} {2007})}\BibitemShut {NoStop}%
\bibitem [{\citenamefont {Li}\ and\ \citenamefont
  {Pendry}(2008)}]{li2008hiding}%
  \BibitemOpen
  \bibfield  {author} {\bibinfo {author} {\bibfnamefont {J.}~\bibnamefont
  {Li}}\ and\ \bibinfo {author} {\bibfnamefont {J.~B.}\ \bibnamefont
  {Pendry}},\ }\bibfield  {title} {\enquote {\bibinfo {title} {Hiding under the
  carpet: a new strategy for cloaking},}\ }\href@noop {} {\bibfield  {journal}
  {\bibinfo  {journal} {Physical {R}eview {L}etters}\ }\textbf {\bibinfo
  {volume} {101}},\ \bibinfo {pages} {203901} (\bibinfo {year}
  {2008})}\BibitemShut {NoStop}%
\bibitem [{\citenamefont {Valentine}\ \emph {et~al.}(2009)\citenamefont
  {Valentine}, \citenamefont {Li}, \citenamefont {Zentgraf}, \citenamefont
  {Bartal},\ and\ \citenamefont {Zhang}}]{valentine2009optical}%
  \BibitemOpen
  \bibfield  {author} {\bibinfo {author} {\bibfnamefont {J.}~\bibnamefont
  {Valentine}}, \bibinfo {author} {\bibfnamefont {J.}~\bibnamefont {Li}},
  \bibinfo {author} {\bibfnamefont {T.}~\bibnamefont {Zentgraf}}, \bibinfo
  {author} {\bibfnamefont {G.}~\bibnamefont {Bartal}}, \ and\ \bibinfo {author}
  {\bibfnamefont {X.}~\bibnamefont {Zhang}},\ }\bibfield  {title} {\enquote
  {\bibinfo {title} {An optical cloak made of dielectrics},}\ }\href@noop {}
  {\bibfield  {journal} {\bibinfo  {journal} {Nature Materials}\ }\textbf
  {\bibinfo {volume} {8}},\ \bibinfo {pages} {568--571} (\bibinfo {year}
  {2009})}\BibitemShut {NoStop}%
\bibitem [{\citenamefont {Al{\`u}}(2009)}]{alu2009mantle}%
  \BibitemOpen
  \bibfield  {author} {\bibinfo {author} {\bibfnamefont {A.}~\bibnamefont
  {Al{\`u}}},\ }\bibfield  {title} {\enquote {\bibinfo {title} {Mantle cloak:
  Invisibility induced by a surface},}\ }\href@noop {} {\bibfield  {journal}
  {\bibinfo  {journal} {Physical Review B}\ }\textbf {\bibinfo {volume} {80}},\
  \bibinfo {pages} {245115} (\bibinfo {year} {2009})}\BibitemShut {NoStop}%
\bibitem [{\citenamefont {Popa}, \citenamefont {Zigoneanu},\ and\ \citenamefont
  {Cummer}(2011)}]{popa2011experimental}%
  \BibitemOpen
  \bibfield  {author} {\bibinfo {author} {\bibfnamefont {B.-I.}\ \bibnamefont
  {Popa}}, \bibinfo {author} {\bibfnamefont {L.}~\bibnamefont {Zigoneanu}}, \
  and\ \bibinfo {author} {\bibfnamefont {S.~A.}\ \bibnamefont {Cummer}},\
  }\bibfield  {title} {\enquote {\bibinfo {title} {Experimental acoustic ground
  cloak in air},}\ }\href@noop {} {\bibfield  {journal} {\bibinfo  {journal}
  {Physical {R}eview {L}etters}\ }\textbf {\bibinfo {volume} {106}},\ \bibinfo
  {pages} {253901} (\bibinfo {year} {2011})}\BibitemShut {NoStop}%
\bibitem [{\citenamefont {Zhu}\ \emph {et~al.}(2013)\citenamefont {Zhu},
  \citenamefont {Feng}, \citenamefont {Zhang}, \citenamefont {Yin},\ and\
  \citenamefont {Zhang}}]{zhu2013one}%
  \BibitemOpen
  \bibfield  {author} {\bibinfo {author} {\bibfnamefont {X.}~\bibnamefont
  {Zhu}}, \bibinfo {author} {\bibfnamefont {L.}~\bibnamefont {Feng}}, \bibinfo
  {author} {\bibfnamefont {P.}~\bibnamefont {Zhang}}, \bibinfo {author}
  {\bibfnamefont {X.}~\bibnamefont {Yin}}, \ and\ \bibinfo {author}
  {\bibfnamefont {X.}~\bibnamefont {Zhang}},\ }\bibfield  {title} {\enquote
  {\bibinfo {title} {One-way invisible cloak using parity-time symmetric
  transformation optics},}\ }\href@noop {} {\bibfield  {journal} {\bibinfo
  {journal} {Optics Letters}\ }\textbf {\bibinfo {volume} {38}},\ \bibinfo
  {pages} {2821--2824} (\bibinfo {year} {2013})}\BibitemShut {NoStop}%
\bibitem [{\citenamefont {Zigoneanu}, \citenamefont {Popa},\ and\ \citenamefont
  {Cummer}(2014)}]{zigoneanu2014three}%
  \BibitemOpen
  \bibfield  {author} {\bibinfo {author} {\bibfnamefont {L.}~\bibnamefont
  {Zigoneanu}}, \bibinfo {author} {\bibfnamefont {B.-I.}\ \bibnamefont {Popa}},
  \ and\ \bibinfo {author} {\bibfnamefont {S.~A.}\ \bibnamefont {Cummer}},\
  }\bibfield  {title} {\enquote {\bibinfo {title} {Three-dimensional broadband
  omnidirectional acoustic ground cloak},}\ }\href@noop {} {\bibfield
  {journal} {\bibinfo  {journal} {Nature {M}aterials}\ }\textbf {\bibinfo
  {volume} {13}},\ \bibinfo {pages} {352--355} (\bibinfo {year}
  {2014})}\BibitemShut {NoStop}%
\bibitem [{\citenamefont {Fan}\ \emph {et~al.}(2020)\citenamefont {Fan},
  \citenamefont {Zhao}, \citenamefont {Cao}, \citenamefont {Zhu}, \citenamefont
  {Chen}, \citenamefont {Wang}, \citenamefont {Donda}, \citenamefont {Wang},\
  and\ \citenamefont {Assouar}}]{fan2020reconfigurable}%
  \BibitemOpen
  \bibfield  {author} {\bibinfo {author} {\bibfnamefont {S.-W.}\ \bibnamefont
  {Fan}}, \bibinfo {author} {\bibfnamefont {S.-D.}\ \bibnamefont {Zhao}},
  \bibinfo {author} {\bibfnamefont {L.}~\bibnamefont {Cao}}, \bibinfo {author}
  {\bibfnamefont {Y.}~\bibnamefont {Zhu}}, \bibinfo {author} {\bibfnamefont
  {A.-L.}\ \bibnamefont {Chen}}, \bibinfo {author} {\bibfnamefont {Y.-F.}\
  \bibnamefont {Wang}}, \bibinfo {author} {\bibfnamefont {K.}~\bibnamefont
  {Donda}}, \bibinfo {author} {\bibfnamefont {Y.-S.}\ \bibnamefont {Wang}}, \
  and\ \bibinfo {author} {\bibfnamefont {B.}~\bibnamefont {Assouar}},\
  }\bibfield  {title} {\enquote {\bibinfo {title} {Reconfigurable curved
  metasurface for acoustic cloaking and illusion},}\ }\href@noop {} {\bibfield
  {journal} {\bibinfo  {journal} {Physical Review B}\ }\textbf {\bibinfo
  {volume} {101}},\ \bibinfo {pages} {024104} (\bibinfo {year}
  {2020})}\BibitemShut {NoStop}%
\bibitem [{\citenamefont {Chen}, \citenamefont {Chan},\ and\ \citenamefont
  {Sheng}(2010)}]{chen2010transformation}%
  \BibitemOpen
  \bibfield  {author} {\bibinfo {author} {\bibfnamefont {H.}~\bibnamefont
  {Chen}}, \bibinfo {author} {\bibfnamefont {C.~T.}\ \bibnamefont {Chan}}, \
  and\ \bibinfo {author} {\bibfnamefont {P.}~\bibnamefont {Sheng}},\ }\bibfield
   {title} {\enquote {\bibinfo {title} {Transformation optics and
  metamaterials},}\ }\href@noop {} {\bibfield  {journal} {\bibinfo  {journal}
  {Nature Materials}\ }\textbf {\bibinfo {volume} {9}},\ \bibinfo {pages}
  {387--396} (\bibinfo {year} {2010})}\BibitemShut {NoStop}%
\bibitem [{\citenamefont {Zhang}, \citenamefont {Pendry},\ and\ \citenamefont
  {Luo}(2019)}]{zhang2019transformation}%
  \BibitemOpen
  \bibfield  {author} {\bibinfo {author} {\bibfnamefont {J.}~\bibnamefont
  {Zhang}}, \bibinfo {author} {\bibfnamefont {J.~B.}\ \bibnamefont {Pendry}}, \
  and\ \bibinfo {author} {\bibfnamefont {Y.}~\bibnamefont {Luo}},\ }\bibfield
  {title} {\enquote {\bibinfo {title} {Transformation optics from macroscopic
  to nanoscale regimes: a review},}\ }\href@noop {} {\bibfield  {journal}
  {\bibinfo  {journal} {Advanced Photonics}\ }\textbf {\bibinfo {volume} {1}},\
  \bibinfo {pages} {014001--014001} (\bibinfo {year} {2019})}\BibitemShut
  {NoStop}%
\bibitem [{\citenamefont {Fridman}\ \emph {et~al.}(2012)\citenamefont
  {Fridman}, \citenamefont {Farsi}, \citenamefont {Okawachi},\ and\
  \citenamefont {Gaeta}}]{fridman2012demonstration}%
  \BibitemOpen
  \bibfield  {author} {\bibinfo {author} {\bibfnamefont {M.}~\bibnamefont
  {Fridman}}, \bibinfo {author} {\bibfnamefont {A.}~\bibnamefont {Farsi}},
  \bibinfo {author} {\bibfnamefont {Y.}~\bibnamefont {Okawachi}}, \ and\
  \bibinfo {author} {\bibfnamefont {A.~L.}\ \bibnamefont {Gaeta}},\ }\bibfield
  {title} {\enquote {\bibinfo {title} {Demonstration of temporal cloaking},}\
  }\href@noop {} {\bibfield  {journal} {\bibinfo  {journal} {Nature}\ }\textbf
  {\bibinfo {volume} {481}},\ \bibinfo {pages} {62--65} (\bibinfo {year}
  {2012})}\BibitemShut {NoStop}%
\bibitem [{\citenamefont {Lukens}, \citenamefont {Leaird},\ and\ \citenamefont
  {Weiner}(2013)}]{lukens2013temporal}%
  \BibitemOpen
  \bibfield  {author} {\bibinfo {author} {\bibfnamefont {J.~M.}\ \bibnamefont
  {Lukens}}, \bibinfo {author} {\bibfnamefont {D.~E.}\ \bibnamefont {Leaird}},
  \ and\ \bibinfo {author} {\bibfnamefont {A.~M.}\ \bibnamefont {Weiner}},\
  }\bibfield  {title} {\enquote {\bibinfo {title} {A temporal cloak at
  telecommunication data rate},}\ }\href@noop {} {\bibfield  {journal}
  {\bibinfo  {journal} {Nature}\ }\textbf {\bibinfo {volume} {498}},\ \bibinfo
  {pages} {205--208} (\bibinfo {year} {2013})}\BibitemShut {NoStop}%
\bibitem [{\citenamefont {Chremmos}(2014)}]{chremmos2014temporal}%
  \BibitemOpen
  \bibfield  {author} {\bibinfo {author} {\bibfnamefont {I.}~\bibnamefont
  {Chremmos}},\ }\bibfield  {title} {\enquote {\bibinfo {title} {Temporal
  cloaking with accelerating wave packets},}\ }\href@noop {} {\bibfield
  {journal} {\bibinfo  {journal} {Optics Letters}\ }\textbf {\bibinfo {volume}
  {39}},\ \bibinfo {pages} {4611--4614} (\bibinfo {year} {2014})}\BibitemShut
  {NoStop}%
\bibitem [{\citenamefont {Zhou}\ \emph {et~al.}(2017)\citenamefont {Zhou},
  \citenamefont {Dong}, \citenamefont {Yan},\ and\ \citenamefont
  {Yang}}]{zhou2017temporal}%
  \BibitemOpen
  \bibfield  {author} {\bibinfo {author} {\bibfnamefont {F.}~\bibnamefont
  {Zhou}}, \bibinfo {author} {\bibfnamefont {J.}~\bibnamefont {Dong}}, \bibinfo
  {author} {\bibfnamefont {S.}~\bibnamefont {Yan}}, \ and\ \bibinfo {author}
  {\bibfnamefont {T.}~\bibnamefont {Yang}},\ }\bibfield  {title} {\enquote
  {\bibinfo {title} {Temporal cloak with large fractional hiding window at
  telecommunication data rate},}\ }\href@noop {} {\bibfield  {journal}
  {\bibinfo  {journal} {Optics Communications}\ }\textbf {\bibinfo {volume}
  {388}},\ \bibinfo {pages} {77--83} (\bibinfo {year} {2017})}\BibitemShut
  {NoStop}%
\bibitem [{\citenamefont {Li}\ \emph {et~al.}(2017)\citenamefont {Li},
  \citenamefont {Wang}, \citenamefont {Kang}, \citenamefont {Wei},
  \citenamefont {Yung},\ and\ \citenamefont {Wong}}]{li2017extended}%
  \BibitemOpen
  \bibfield  {author} {\bibinfo {author} {\bibfnamefont {B.}~\bibnamefont
  {Li}}, \bibinfo {author} {\bibfnamefont {X.}~\bibnamefont {Wang}}, \bibinfo
  {author} {\bibfnamefont {J.}~\bibnamefont {Kang}}, \bibinfo {author}
  {\bibfnamefont {Y.}~\bibnamefont {Wei}}, \bibinfo {author} {\bibfnamefont
  {T.}~\bibnamefont {Yung}}, \ and\ \bibinfo {author} {\bibfnamefont {K.~K.}\
  \bibnamefont {Wong}},\ }\bibfield  {title} {\enquote {\bibinfo {title}
  {Extended temporal cloak based on the inverse temporal talbot effect},}\
  }\href@noop {} {\bibfield  {journal} {\bibinfo  {journal} {Optics Letters}\
  }\textbf {\bibinfo {volume} {42}},\ \bibinfo {pages} {767--770} (\bibinfo
  {year} {2017})}\BibitemShut {NoStop}%
\bibitem [{\citenamefont {Zhou}\ \emph {et~al.}(2019)\citenamefont {Zhou},
  \citenamefont {Yan}, \citenamefont {Zhou}, \citenamefont {Wang},
  \citenamefont {Qiu}, \citenamefont {Dong}, \citenamefont {Zhou},
  \citenamefont {Ding}, \citenamefont {Qiu},\ and\ \citenamefont
  {Zhang}}]{zhou2019field}%
  \BibitemOpen
  \bibfield  {author} {\bibinfo {author} {\bibfnamefont {F.}~\bibnamefont
  {Zhou}}, \bibinfo {author} {\bibfnamefont {S.}~\bibnamefont {Yan}}, \bibinfo
  {author} {\bibfnamefont {H.}~\bibnamefont {Zhou}}, \bibinfo {author}
  {\bibfnamefont {X.}~\bibnamefont {Wang}}, \bibinfo {author} {\bibfnamefont
  {H.}~\bibnamefont {Qiu}}, \bibinfo {author} {\bibfnamefont {J.}~\bibnamefont
  {Dong}}, \bibinfo {author} {\bibfnamefont {L.}~\bibnamefont {Zhou}}, \bibinfo
  {author} {\bibfnamefont {Y.}~\bibnamefont {Ding}}, \bibinfo {author}
  {\bibfnamefont {C.-W.}\ \bibnamefont {Qiu}}, \ and\ \bibinfo {author}
  {\bibfnamefont {X.}~\bibnamefont {Zhang}},\ }\bibfield  {title} {\enquote
  {\bibinfo {title} {Field-programmable silicon temporal cloak},}\ }\href@noop
  {} {\bibfield  {journal} {\bibinfo  {journal} {Nature Communications}\
  }\textbf {\bibinfo {volume} {10}},\ \bibinfo {pages} {2726} (\bibinfo {year}
  {2019})}\BibitemShut {NoStop}%
\bibitem [{\citenamefont {McCall}\ \emph {et~al.}(2010)\citenamefont {McCall},
  \citenamefont {Favaro}, \citenamefont {Kinsler},\ and\ \citenamefont
  {Boardman}}]{mccall2010spacetime}%
  \BibitemOpen
  \bibfield  {author} {\bibinfo {author} {\bibfnamefont {M.~W.}\ \bibnamefont
  {McCall}}, \bibinfo {author} {\bibfnamefont {A.}~\bibnamefont {Favaro}},
  \bibinfo {author} {\bibfnamefont {P.}~\bibnamefont {Kinsler}}, \ and\
  \bibinfo {author} {\bibfnamefont {A.}~\bibnamefont {Boardman}},\ }\bibfield
  {title} {\enquote {\bibinfo {title} {A spacetime cloak, or a history
  editor},}\ }\href@noop {} {\bibfield  {journal} {\bibinfo  {journal} {Journal
  of Optics}\ }\textbf {\bibinfo {volume} {13}},\ \bibinfo {pages} {024003}
  (\bibinfo {year} {2010})}\BibitemShut {NoStop}%
\bibitem [{\citenamefont {Kinsler}\ and\ \citenamefont
  {McCall}(2014)}]{kinsler2014cloaks}%
  \BibitemOpen
  \bibfield  {author} {\bibinfo {author} {\bibfnamefont {P.}~\bibnamefont
  {Kinsler}}\ and\ \bibinfo {author} {\bibfnamefont {M.~W.}\ \bibnamefont
  {McCall}},\ }\bibfield  {title} {\enquote {\bibinfo {title} {Cloaks, editors,
  and bubbles: applications of spacetime transformation theory},}\ }\href@noop
  {} {\bibfield  {journal} {\bibinfo  {journal} {Annalen der Physik}\ }\textbf
  {\bibinfo {volume} {526}},\ \bibinfo {pages} {51--62} (\bibinfo {year}
  {2014})}\BibitemShut {NoStop}%
\bibitem [{\citenamefont {Akl}\ and\ \citenamefont {Baz}(2010)}]{akl2010multi}%
  \BibitemOpen
  \bibfield  {author} {\bibinfo {author} {\bibfnamefont {W.}~\bibnamefont
  {Akl}}\ and\ \bibinfo {author} {\bibfnamefont {A.}~\bibnamefont {Baz}},\
  }\bibfield  {title} {\enquote {\bibinfo {title} {Multi-cell active acoustic
  metamaterial with programmable bulk modulus},}\ }\href@noop {} {\bibfield
  {journal} {\bibinfo  {journal} {Journal of Intelligent Material Systems and
  Structures}\ }\textbf {\bibinfo {volume} {21}},\ \bibinfo {pages} {541--556}
  (\bibinfo {year} {2010})}\BibitemShut {NoStop}%
\bibitem [{\citenamefont {Popa}\ \emph {et~al.}(2015)\citenamefont {Popa},
  \citenamefont {Shinde}, \citenamefont {Konneker},\ and\ \citenamefont
  {Cummer}}]{popa2015active}%
  \BibitemOpen
  \bibfield  {author} {\bibinfo {author} {\bibfnamefont {B.-I.}\ \bibnamefont
  {Popa}}, \bibinfo {author} {\bibfnamefont {D.}~\bibnamefont {Shinde}},
  \bibinfo {author} {\bibfnamefont {A.}~\bibnamefont {Konneker}}, \ and\
  \bibinfo {author} {\bibfnamefont {S.~A.}\ \bibnamefont {Cummer}},\ }\bibfield
   {title} {\enquote {\bibinfo {title} {Active acoustic metamaterials
  reconfigurable in real time},}\ }\href@noop {} {\bibfield  {journal}
  {\bibinfo  {journal} {Physical Review B}\ }\textbf {\bibinfo {volume} {91}},\
  \bibinfo {pages} {220303} (\bibinfo {year} {2015})}\BibitemShut {NoStop}%
\bibitem [{\citenamefont {B{\"o}rsing}\ \emph {et~al.}(2019)\citenamefont
  {B{\"o}rsing}, \citenamefont {Becker}, \citenamefont {Curtis}, \citenamefont
  {van Manen}, \citenamefont {Haag},\ and\ \citenamefont
  {Robertsson}}]{borsing2019cloaking}%
  \BibitemOpen
  \bibfield  {author} {\bibinfo {author} {\bibfnamefont {N.}~\bibnamefont
  {B{\"o}rsing}}, \bibinfo {author} {\bibfnamefont {T.~S.}\ \bibnamefont
  {Becker}}, \bibinfo {author} {\bibfnamefont {A.}~\bibnamefont {Curtis}},
  \bibinfo {author} {\bibfnamefont {D.-J.}\ \bibnamefont {van Manen}}, \bibinfo
  {author} {\bibfnamefont {T.}~\bibnamefont {Haag}}, \ and\ \bibinfo {author}
  {\bibfnamefont {J.~O.}\ \bibnamefont {Robertsson}},\ }\bibfield  {title}
  {\enquote {\bibinfo {title} {Cloaking and holography experiments using
  immersive boundary conditions},}\ }\href@noop {} {\bibfield  {journal}
  {\bibinfo  {journal} {Physical Review Applied}\ }\textbf {\bibinfo {volume}
  {12}},\ \bibinfo {pages} {024011} (\bibinfo {year} {2019})}\BibitemShut
  {NoStop}%
\bibitem [{\citenamefont {Hofmann}\ \emph {et~al.}(2019)\citenamefont
  {Hofmann}, \citenamefont {Helbig}, \citenamefont {Lee}, \citenamefont
  {Greiter},\ and\ \citenamefont {Thomale}}]{hofmann2019chiral}%
  \BibitemOpen
  \bibfield  {author} {\bibinfo {author} {\bibfnamefont {T.}~\bibnamefont
  {Hofmann}}, \bibinfo {author} {\bibfnamefont {T.}~\bibnamefont {Helbig}},
  \bibinfo {author} {\bibfnamefont {C.~H.}\ \bibnamefont {Lee}}, \bibinfo
  {author} {\bibfnamefont {M.}~\bibnamefont {Greiter}}, \ and\ \bibinfo
  {author} {\bibfnamefont {R.}~\bibnamefont {Thomale}},\ }\bibfield  {title}
  {\enquote {\bibinfo {title} {Chiral voltage propagation and calibration in a
  topolectrical {C}hern circuit},}\ }\href@noop {} {\bibfield  {journal}
  {\bibinfo  {journal} {Physical Review Letters}\ }\textbf {\bibinfo {volume}
  {122}},\ \bibinfo {pages} {247702} (\bibinfo {year} {2019})}\BibitemShut
  {NoStop}%
\bibitem [{\citenamefont {Sirota}, \citenamefont {Semperlotti},\ and\
  \citenamefont {Annaswamy}(2019)}]{sirota2019tunable}%
  \BibitemOpen
  \bibfield  {author} {\bibinfo {author} {\bibfnamefont {L.}~\bibnamefont
  {Sirota}}, \bibinfo {author} {\bibfnamefont {F.}~\bibnamefont {Semperlotti}},
  \ and\ \bibinfo {author} {\bibfnamefont {A.~M.}\ \bibnamefont {Annaswamy}},\
  }\bibfield  {title} {\enquote {\bibinfo {title} {Tunable and reconfigurable
  mechanical transmission-line metamaterials via direct active feedback
  control},}\ }\href@noop {} {\bibfield  {journal} {\bibinfo  {journal}
  {Mechanical Systems and Signal Processing}\ }\textbf {\bibinfo {volume}
  {123}},\ \bibinfo {pages} {117--130} (\bibinfo {year} {2019})}\BibitemShut
  {NoStop}%
\bibitem [{\citenamefont {Sirota}\ and\ \citenamefont
  {Annaswamy}(2019)}]{sirota2019active}%
  \BibitemOpen
  \bibfield  {author} {\bibinfo {author} {\bibfnamefont {L.}~\bibnamefont
  {Sirota}}\ and\ \bibinfo {author} {\bibfnamefont {A.~M.}\ \bibnamefont
  {Annaswamy}},\ }\bibfield  {title} {\enquote {\bibinfo {title} {Active wave
  suppression in the interior of a one-dimensional domain},}\ }\href@noop {}
  {\bibfield  {journal} {\bibinfo  {journal} {Automatica}\ }\textbf {\bibinfo
  {volume} {100}},\ \bibinfo {pages} {403--406} (\bibinfo {year}
  {2019})}\BibitemShut {NoStop}%
\bibitem [{\citenamefont {Sirota}\ and\ \citenamefont
  {Annaswamy}(2020)}]{sirota2020modeling}%
  \BibitemOpen
  \bibfield  {author} {\bibinfo {author} {\bibfnamefont {L.}~\bibnamefont
  {Sirota}}\ and\ \bibinfo {author} {\bibfnamefont {A.~M.}\ \bibnamefont
  {Annaswamy}},\ }\bibfield  {title} {\enquote {\bibinfo {title} {Active
  boundary and interior absorbers for one-dimensional wave propagation:
  {A}pplication to transmission-line metamaterials},}\ }\href@noop {}
  {\bibfield  {journal} {\bibinfo  {journal} {Automatica}\ }\textbf {\bibinfo
  {volume} {117}},\ \bibinfo {pages} {108--855} (\bibinfo {year}
  {2020})}\BibitemShut {NoStop}%
\bibitem [{\citenamefont {Becker}\ \emph {et~al.}(2020)\citenamefont {Becker},
  \citenamefont {B{\"o}rsing}, \citenamefont {Haag}, \citenamefont
  {B{\"a}rlocher}, \citenamefont {Donahue}, \citenamefont {Curtis},
  \citenamefont {Robertsson},\ and\ \citenamefont {van
  Manen}}]{becker2020real}%
  \BibitemOpen
  \bibfield  {author} {\bibinfo {author} {\bibfnamefont {T.~S.}\ \bibnamefont
  {Becker}}, \bibinfo {author} {\bibfnamefont {N.}~\bibnamefont {B{\"o}rsing}},
  \bibinfo {author} {\bibfnamefont {T.}~\bibnamefont {Haag}}, \bibinfo {author}
  {\bibfnamefont {C.}~\bibnamefont {B{\"a}rlocher}}, \bibinfo {author}
  {\bibfnamefont {C.~M.}\ \bibnamefont {Donahue}}, \bibinfo {author}
  {\bibfnamefont {A.}~\bibnamefont {Curtis}}, \bibinfo {author} {\bibfnamefont
  {J.~O.}\ \bibnamefont {Robertsson}}, \ and\ \bibinfo {author} {\bibfnamefont
  {D.-J.}\ \bibnamefont {van Manen}},\ }\bibfield  {title} {\enquote {\bibinfo
  {title} {Real-time immersion of physical experiments in virtual wave-physics
  domains},}\ }\href@noop {} {\bibfield  {journal} {\bibinfo  {journal}
  {Physical Review Applied}\ }\textbf {\bibinfo {volume} {13}},\ \bibinfo
  {pages} {064061} (\bibinfo {year} {2020})}\BibitemShut {NoStop}%
\bibitem [{\citenamefont {Scheibner}, \citenamefont {Irvine},\ and\
  \citenamefont {Vitelli}(2020)}]{scheibner2020non}%
  \BibitemOpen
  \bibfield  {author} {\bibinfo {author} {\bibfnamefont {C.}~\bibnamefont
  {Scheibner}}, \bibinfo {author} {\bibfnamefont {W.~T.}\ \bibnamefont
  {Irvine}}, \ and\ \bibinfo {author} {\bibfnamefont {V.}~\bibnamefont
  {Vitelli}},\ }\bibfield  {title} {\enquote {\bibinfo {title} {Non-{H}ermitian
  band topology and skin modes in active elastic media},}\ }\href@noop {}
  {\bibfield  {journal} {\bibinfo  {journal} {Physical {R}eview {L}etters}\
  }\textbf {\bibinfo {volume} {125}},\ \bibinfo {pages} {118001} (\bibinfo
  {year} {2020})}\BibitemShut {NoStop}%
\bibitem [{\citenamefont {Rosa}\ and\ \citenamefont
  {Ruzzene}(2020)}]{rosa2020dynamics}%
  \BibitemOpen
  \bibfield  {author} {\bibinfo {author} {\bibfnamefont {M.~I.}\ \bibnamefont
  {Rosa}}\ and\ \bibinfo {author} {\bibfnamefont {M.}~\bibnamefont {Ruzzene}},\
  }\bibfield  {title} {\enquote {\bibinfo {title} {Dynamics and topology of
  non-{H}ermitian elastic lattices with non-local feedback control
  interactions},}\ }\href@noop {} {\bibfield  {journal} {\bibinfo  {journal}
  {New {J}ournal of {P}hysics}\ }\textbf {\bibinfo {volume} {22}},\ \bibinfo
  {pages} {053004} (\bibinfo {year} {2020})}\BibitemShut {NoStop}%
\bibitem [{\citenamefont {Cho}\ \emph {et~al.}(2020)\citenamefont {Cho},
  \citenamefont {Wen}, \citenamefont {Park},\ and\ \citenamefont
  {Li}}]{cho2020digitally}%
  \BibitemOpen
  \bibfield  {author} {\bibinfo {author} {\bibfnamefont {C.}~\bibnamefont
  {Cho}}, \bibinfo {author} {\bibfnamefont {X.}~\bibnamefont {Wen}}, \bibinfo
  {author} {\bibfnamefont {N.}~\bibnamefont {Park}}, \ and\ \bibinfo {author}
  {\bibfnamefont {J.}~\bibnamefont {Li}},\ }\bibfield  {title} {\enquote
  {\bibinfo {title} {Digitally virtualized atoms for acoustic metamaterials},}\
  }\href@noop {} {\bibfield  {journal} {\bibinfo  {journal} {Nature
  Communications}\ }\textbf {\bibinfo {volume} {11}},\ \bibinfo {pages} {1--8}
  (\bibinfo {year} {2020})}\BibitemShut {NoStop}%
\bibitem [{\citenamefont {Ghatak}\ \emph {et~al.}(2020)\citenamefont {Ghatak},
  \citenamefont {Brandenbourger}, \citenamefont {van Wezel},\ and\
  \citenamefont {Coulais}}]{ghatak2020observation}%
  \BibitemOpen
  \bibfield  {author} {\bibinfo {author} {\bibfnamefont {A.}~\bibnamefont
  {Ghatak}}, \bibinfo {author} {\bibfnamefont {M.}~\bibnamefont
  {Brandenbourger}}, \bibinfo {author} {\bibfnamefont {J.}~\bibnamefont {van
  Wezel}}, \ and\ \bibinfo {author} {\bibfnamefont {C.}~\bibnamefont
  {Coulais}},\ }\bibfield  {title} {\enquote {\bibinfo {title} {Observation of
  non-{H}ermitian topology and its bulk-edge correspondence in an active
  mechanical metamaterial},}\ }\href@noop {} {\bibfield  {journal} {\bibinfo
  {journal} {Proceedings of the {N}ational {A}cademy of {S}ciences}\ }
  (\bibinfo {year} {2020})}\BibitemShut {NoStop}%
\bibitem [{\citenamefont {Kotwal}\ \emph {et~al.}(2021)\citenamefont {Kotwal},
  \citenamefont {Moseley}, \citenamefont {Stegmaier}, \citenamefont {Imhof},
  \citenamefont {Brand}, \citenamefont {Kie{\ss}ling}, \citenamefont {Thomale},
  \citenamefont {Ronellenfitsch},\ and\ \citenamefont
  {Dunkel}}]{kotwal2021active}%
  \BibitemOpen
  \bibfield  {author} {\bibinfo {author} {\bibfnamefont {T.}~\bibnamefont
  {Kotwal}}, \bibinfo {author} {\bibfnamefont {F.}~\bibnamefont {Moseley}},
  \bibinfo {author} {\bibfnamefont {A.}~\bibnamefont {Stegmaier}}, \bibinfo
  {author} {\bibfnamefont {S.}~\bibnamefont {Imhof}}, \bibinfo {author}
  {\bibfnamefont {H.}~\bibnamefont {Brand}}, \bibinfo {author} {\bibfnamefont
  {T.}~\bibnamefont {Kie{\ss}ling}}, \bibinfo {author} {\bibfnamefont
  {R.}~\bibnamefont {Thomale}}, \bibinfo {author} {\bibfnamefont
  {H.}~\bibnamefont {Ronellenfitsch}}, \ and\ \bibinfo {author} {\bibfnamefont
  {J.}~\bibnamefont {Dunkel}},\ }\bibfield  {title} {\enquote {\bibinfo {title}
  {Active topolectrical circuits},}\ }\href@noop {} {\bibfield  {journal}
  {\bibinfo  {journal} {Proceedings of the National Academy of Sciences}\
  }\textbf {\bibinfo {volume} {118}} (\bibinfo {year} {2021})}\BibitemShut
  {NoStop}%
\bibitem [{\citenamefont {You}\ \emph {et~al.}(2021)\citenamefont {You},
  \citenamefont {Ma}, \citenamefont {Lan}, \citenamefont {Xiao}, \citenamefont
  {Panoiu},\ and\ \citenamefont {Cui}}]{you2021reprogrammable}%
  \BibitemOpen
  \bibfield  {author} {\bibinfo {author} {\bibfnamefont {J.~W.}\ \bibnamefont
  {You}}, \bibinfo {author} {\bibfnamefont {Q.}~\bibnamefont {Ma}}, \bibinfo
  {author} {\bibfnamefont {Z.}~\bibnamefont {Lan}}, \bibinfo {author}
  {\bibfnamefont {Q.}~\bibnamefont {Xiao}}, \bibinfo {author} {\bibfnamefont
  {N.~C.}\ \bibnamefont {Panoiu}}, \ and\ \bibinfo {author} {\bibfnamefont
  {T.~J.}\ \bibnamefont {Cui}},\ }\bibfield  {title} {\enquote {\bibinfo
  {title} {Reprogrammable plasmonic topological insulators with ultrafast
  control},}\ }\href@noop {} {\bibfield  {journal} {\bibinfo  {journal} {Nature
  Communications}\ }\textbf {\bibinfo {volume} {12}},\ \bibinfo {pages} {1--7}
  (\bibinfo {year} {2021})}\BibitemShut {NoStop}%
\bibitem [{\citenamefont {Geib}\ \emph {et~al.}(2021)\citenamefont {Geib},
  \citenamefont {Sasmal}, \citenamefont {Wang}, \citenamefont {Zhai},
  \citenamefont {Popa},\ and\ \citenamefont {Grosh}}]{geib2021tunable}%
  \BibitemOpen
  \bibfield  {author} {\bibinfo {author} {\bibfnamefont {N.}~\bibnamefont
  {Geib}}, \bibinfo {author} {\bibfnamefont {A.}~\bibnamefont {Sasmal}},
  \bibinfo {author} {\bibfnamefont {Z.}~\bibnamefont {Wang}}, \bibinfo {author}
  {\bibfnamefont {Y.}~\bibnamefont {Zhai}}, \bibinfo {author} {\bibfnamefont
  {B.-I.}\ \bibnamefont {Popa}}, \ and\ \bibinfo {author} {\bibfnamefont
  {K.}~\bibnamefont {Grosh}},\ }\bibfield  {title} {\enquote {\bibinfo {title}
  {Tunable nonlocal purely active nonreciprocal acoustic media},}\ }\href@noop
  {} {\bibfield  {journal} {\bibinfo  {journal} {Physical Review B}\ }\textbf
  {\bibinfo {volume} {103}},\ \bibinfo {pages} {165427} (\bibinfo {year}
  {2021})}\BibitemShut {NoStop}%
\bibitem [{\citenamefont {Li}\ \emph {et~al.}(2021)\citenamefont {Li},
  \citenamefont {Chen}, \citenamefont {Zhu},\ and\ \citenamefont
  {Huang}}]{li2021active}%
  \BibitemOpen
  \bibfield  {author} {\bibinfo {author} {\bibfnamefont {X.}~\bibnamefont
  {Li}}, \bibinfo {author} {\bibfnamefont {Y.}~\bibnamefont {Chen}}, \bibinfo
  {author} {\bibfnamefont {R.}~\bibnamefont {Zhu}}, \ and\ \bibinfo {author}
  {\bibfnamefont {G.}~\bibnamefont {Huang}},\ }\bibfield  {title} {\enquote
  {\bibinfo {title} {An active meta-layer for optimal flexural wave absorption
  and cloaking},}\ }\href@noop {} {\bibfield  {journal} {\bibinfo  {journal}
  {Mechanical Systems and Signal Processing}\ }\textbf {\bibinfo {volume}
  {149}},\ \bibinfo {pages} {107324} (\bibinfo {year} {2021})}\BibitemShut
  {NoStop}%
\bibitem [{\citenamefont {Stojanoska}\ and\ \citenamefont
  {Shen}(2022)}]{stojanoska2022non}%
  \BibitemOpen
  \bibfield  {author} {\bibinfo {author} {\bibfnamefont {K.}~\bibnamefont
  {Stojanoska}}\ and\ \bibinfo {author} {\bibfnamefont {C.}~\bibnamefont
  {Shen}},\ }\bibfield  {title} {\enquote {\bibinfo {title} {Non-{H}ermitian
  planar elastic metasurface for unidirectional focusing of flexural waves},}\
  }\href@noop {} {\bibfield  {journal} {\bibinfo  {journal} {Applied Physics
  Letters}\ }\textbf {\bibinfo {volume} {120}},\ \bibinfo {pages} {241701}
  (\bibinfo {year} {2022})}\BibitemShut {NoStop}%
\bibitem [{\citenamefont {Cornelius}, \citenamefont {Liu},\ and\ \citenamefont
  {Brio}(2014)}]{cornelius2014finite}%
  \BibitemOpen
  \bibfield  {author} {\bibinfo {author} {\bibfnamefont {J.}~\bibnamefont
  {Cornelius}}, \bibinfo {author} {\bibfnamefont {J.}~\bibnamefont {Liu}}, \
  and\ \bibinfo {author} {\bibfnamefont {M.}~\bibnamefont {Brio}},\ }\bibfield
  {title} {\enquote {\bibinfo {title} {Finite-difference time-domain simulation
  of spacetime cloak},}\ }\href@noop {} {\bibfield  {journal} {\bibinfo
  {journal} {Optics Express}\ }\textbf {\bibinfo {volume} {22}},\ \bibinfo
  {pages} {12087--12095} (\bibinfo {year} {2014})}\BibitemShut {NoStop}%
\end{thebibliography}%

\end{document}